\documentclass[final,letterpaper,12pt]{elsarticle}
\usepackage[centering,top=1in,bottom=1in,left=1in,right=1in]{geometry}
\usepackage[a-1b]{pdfx}   

\usepackage[version=3]{mhchem} 
\usepackage{bm}
\usepackage{mathtools}
\usepackage[]{upgreek}
\usepackage[]{MnSymbol} 
\usepackage[]{gensymb}
\usepackage{adjustbox} 
\usepackage{lscape}
\usepackage{comment}
\usepackage{gensymb}
\usepackage{float}
\usepackage{graphicx}
\usepackage{caption}
\usepackage{subcaption}
\usepackage{amsfonts}
\usepackage{comment}
\usepackage{enumitem}
\usepackage{amsmath}

\usepackage[version=3]{mhchem} 
\usepackage[utf8]{inputenc}
\usepackage{setspace}

\newcommand{\mbs}[1]{\mathbf{#1}}

\newcommand{\mbg}[1]{\boldsymbol{#1}}

\newcommand{\widebar}[1]{\,\overline{\!{#1}}}

\newcommand{\be}{{\mbs{e}}}
\newcommand{\bA}{{\mbs{A}}}
\newcommand{\bB}{{\mbs{B}}}
\newcommand{\bC}{{\mbs{C}}}
\newcommand{\bD}{{\mbs{D}}}
\newcommand{\bF}{{\mbs{F}}}
\newcommand{\bI}{{\mbs{I}}}
\newcommand{\bL}{{\mbs{L}}}
\newcommand{\bM}{{\mbs{M}}}
\newcommand{\bR}{{\mbs{R}}}

\newcommand{\bW}{{\mbs{W}}}
\newcommand{\bZ}{{\mbs{0}}}

\newcommand{\tr}{{\operatorname{tr} \,}}
\newcommand{\dev}{{\operatorname{dev} \,}}

\newcommand{\Fe}{{\mbs{F}^e}}
\newcommand{\Fp}{{\mbs{F}^p}}
\newcommand{\Rel}{{\mbs{R}^e}}

\newcommand{\Fedot}{{\dot{\bF}^e}}
\newcommand{\Fpdot}{{\dot{\bF}^p}}

\newcommand{\eps}{\varepsilon}

\newcommand{\sig}{{\mbg{\sigma}}}
\newcommand{\del}{{\mbg{\delta}}}
\newcommand{\gam}{{\mbg{\gamma}}}
\newcommand{\so} {{\hat{\sigma}_0}}
\newcommand{\soOc} {{\hat{\sigma}_0^c}}
\newcommand{\soI} {{\hat{\sigma}_0^{\rm init}}}

\makeatletter
\newcommand*{\centerfloat}{%
  \parindent \z@
  \leftskip \z@ \@plus 1fil \@minus \textwidth
  \rightskip\leftskip
  \parfillskip \z@skip}
\makeatother

\makeatletter
\def\ps@pprintTitle{%
\let\@oddhead\@empty
\let\@evenhead\@empty
\let\@oddfoot\@empty
\let\@evenfoot\@oddfoot
}
\makeatother

\makeatletter
\def\ps@pprintTitle{%
\let\@oddhead\@empty
\let\@evenhead\@empty
\let\@evenfoot\@oddfoot
}
\makeatother

\begin{document}

\begin{frontmatter}
    \title{A variational critical-state theory of friction}
    \author[1]{Mary Agajanian}
   \author[1,2]{Nadia Lapusta}
   \author[3]{Anna Pandolfi}
   \author[4,5]{Michael Ortiz}
   
   \address[1]{Department of Mechanical and Civil Engineering, California Institute of
Technology, Pasadena, California, USA}
\address[2]{Seismological Laboratory, California Institute of Technology, Pasadena, California, USA}
\address[3]{Dipartimento di Ingegneria Civile e Ambientale, Politecnico di Milano, Milano, Italy}
\address[4]{Graduate Aerospace Laboratories, California Institute of Technology, Pasadena, California, USA}
\address[5]{Centre Internacional De Mètodes Numèrics a L'Enginyeria (CIMNE), Universitat Politècnica de Catalunya, Barcelona, Spain}

    \begin{abstract}
    Friction plays a fundamental role in many natural processes, including earthquakes, landslides, and volcanic eruptions. Earthquakes occur when highly compressed fault surfaces accumulate large enough shear stresses, causing the faults to move relative to one another, or slip. The slip is accommodated within a thin layer of comminuted granular material -- called fault gouge -- between the fault surfaces. As a result, characterizing the mechanical behavior of fault gouge in response to shear is a major open problem in earthquake source physics. Modeling gouge is complicated by large deformations, inelasticity, rate dependence, and volumetric changes. As such, researchers typically rely on empirical formulations to capture the effective response. Here, we systematically develop a variational, finite-kinematics framework for fault gouge.  We first describe a general theory for a rigid-viscoplastic, pressure-sensitive material, where the plasticity evolution follows from the principle of maximum dissipation. Then, we specialize the governing equations for a Cam-Clay material within a shearing and dilating layer. We rely on convexity considerations and experimental observations from consolidation tests of granular layers to calibrate the model and develop explicit solutions for the rate- and state- dependent response of the model to shear tests under constant compressive normal stress and prescribed shearing rate. To validate the model, we select common rate functions and compare numerical material point tests and theoretical solutions to standard laboratory experiments of shearing granular layers. Lastly, we discuss connections of the model to empirical rate-and-state friction laws.  
    \end{abstract}

\end{frontmatter}

\section{Introduction}

Large earthquakes occur as shear ruptures along pre-existing faults—weak interfaces within the Earth’s crust \cite{scholz2019mechanics}. Repeated sliding of these interfaces produces a finely comminuted granular layer known as fault gouge \cite{Chester1993, Faulkner2003}. At seismogenic depths, the gouge layer is highly compacted and undergoes intense shearing due to the relative tectonic motion of the bounding rock masses. The distribution of slip within the gouge may be either homogeneous or localized along discrete shear bands. Furthermore, grain rearrangement during shear induces macroscopic volumetric changes that directly influence the shear resistance of the layer. Geological evidence indicates that, during large earthquakes, faults may slip by several meters within zones less than a millimeter thick, implying extremely large strains across the gouge layer (e.g.~\cite{Chester1998}). Accurate continuum modeling of the finite-deformation, shearing, and dilating behavior of such granular layers is therefore essential for elucidating the conditions under which earthquakes nucleate and propagate.

The stress and volumetric response of sheared granular layers have been the focus of extensive laboratory investigations~\cite{Byerlee1978, Dieterich1978, Dieterich1979, Ruina1983, marone1998laboratory}. In these experiments, a granular assembly is confined under a prescribed normal stress and sheared under controlled loading protocols at relatively slow shear rates (corresponding to slip rates over the layer of $10^{-8}$ to $10^{-3}$ m/s). The resulting shear resistance is largely cohesionless and frictional. That is, the ratio of shear to normal stress defines an effective friction coefficient, typically in the range of 0.6–0.85~\cite{Byerlee1978}. Under constant shear loading rates, the shear resistance approaches a rate-dependent steady state. In response to abrupt changes in shear rate, the shear stress exhibits a transient, non-monotonic evolution: an instantaneous rate-dependent jump followed by a gradual evolution with slip toward a new steady-state value~\cite{Dieterich1979}. These so-called velocity-step tests are used to interrogate shear resistance in two important regimes -- abrupt changes in shear rates and constant shear rates -- and are key to understanding fault gouge behavior during both earthquake nucleation and seismic slip. 

Sheared granular layers also exhibit pronounced consolidation- and rate-dependent volumetric changes. At low confinement, the material dilates under shear; at higher confinement, it compacts~\cite{Das2019, Borja2013}. The transition between these regimes is governed by the pre-consolidation pressure, i.~e., the maximum past pressure experienced by the material. The volumetric response is further rate-dependent, with more dilation at higher shear rates~\cite{marone1990frictional, Segall1995}. This suggests that, at constant loading rates, the layer attains a rate-dependent steady-state volume that increases with shear rate~\cite{marone1990frictional, Segall1995, mair1999friction}. 

A wide range of theoretical models have been proposed to describe the mechanical response of fault gouge and analog materials. In one class of models, the highly-confined, narrow granular layer is interpreted as a frictional interface, with the shear resistance given by empirical rate-and-state friction laws~\cite{Dieterich1979, Ruina1983, marone1998laboratory}, which have been explained with various micro-physical interpretations (e.~g.~\cite{Niemeijer2007, Hulikal2015, Barbot2022}). Other models of fault gouge and analog materials include inertial rheologies~\cite{Henann2013, Nagy2017, Bilotto2025}, shear-transformation-zone theories~\cite{elbanna2014two}, extended continuum models~\cite{Collins-Craft2020}, and micromechanical formulations~\cite{Vakis2018}. Continuum approaches frequently employ pressure-dependent viscoplastic models such as the Drucker–Prager or Mohr–Coulomb formulations~\cite{egholm2008mechanics, dunham2011earthquake}. To account for hardening at high pressures, some models augment the yield surface with an elliptic cap~\cite{dolarevic2007modified}. Other extensions include the effect of grain crushing~\cite{Einav2007, Tengattini2016}. In most such theories, however, the yield surface is prescribed \textit{a priori}, thereby constraining the possible material response and limiting the adaptability of the formulation to large-strain kinematics.

The long-term bulk mechanical behavior of granular and cohesive-frictional geological materials is often described within the critical-state framework~\cite{roscoe:1968, schofield:1968, ortiz2004variational, Borja2013}. These theories are motivated by the empirical observation that, after sufficient shear, a granular material attains a critical state characterized by constant stress and volume during continued plastic deformation. Critical-state plasticity thereby captures the essential features of pressure dependence, yielding, plastic flow, and dilatancy in granular media.

The Cam–Clay model stands as one of the most influential among the various critical-state formulations. Developed at the University of Cambridge in the 1960s, it provides a constitutive framework that unifies compressibility, yielding, and stress-dependent stiffness in natural clays. Foundational works by Roscoe, Schofield, and Wroth~\cite{roscoe:1958, schofield:1968, burland:1969} established both the experimental basis and theoretical structure of the critical-state concept. The modified Cam–Clay model describes material behavior through an elliptical yield surface in the mean effective stress–deviatoric stress plane, which expands (hardens) or contracts (softens) with plastic volumetric strain. The critical-state line delineates the transition between the hardening and softening responses. Extensions of Cam–Clay to finite deformations have been achieved through the assumption of a multiplicative decomposition of the deformation gradient~\cite{borja:1998, callari:1998, ortiz2004variational, borja:2006}.

In the present study, we adapt the critical-state plasticity framework originally developed for clays and soils to model the response of fault gouge. Our formulation builds upon the variational finite-kinematics framework introduced in~\cite{Ortiz1999} and extended to Cam–Clay plasticity in~\cite{ortiz2004variational}. In contrast to~\cite{ortiz2004variational}, where the incremental formulation led to timestep sensitivity, we employ a continuous-time approach. By assuming a specific form of the flow rule and deriving the yield surface from the principle of maximum plastic dissipation, we obtain a consistent, discretization-independent formulation. Under the assumption of rigid elasticity, we derive closed-form solutions for the stress response of a sheared and compressed granular layer.

The remainder of the paper is organized as follows. In Section~\ref{sec::key_mechanisms}, we summarize the key experimental observations that motivate our modeling framework. Section~\ref{sec::governingEquations} presents the general variational formulation for non-isochoric, rigid-plastic materials and specializes it to Cam–Clay plasticity. Section~\ref{sec::shear_layer} further develops the theory for a shearing and dilating granular layer, discusses model calibration, and provides closed-form expressions for simple shear. In Section~\ref{sec::results}, we present numerical results and compare the model predictions with experimental data and existing theoretical models. Finally, Section~\ref{sec:conclusions} summarizes the main findings and outlines directions for future research.

\section{Key mechanisms}
\label{sec::key_mechanisms}
\begin{figure}[H]
    \centering
    \includegraphics[width=0.85\linewidth]{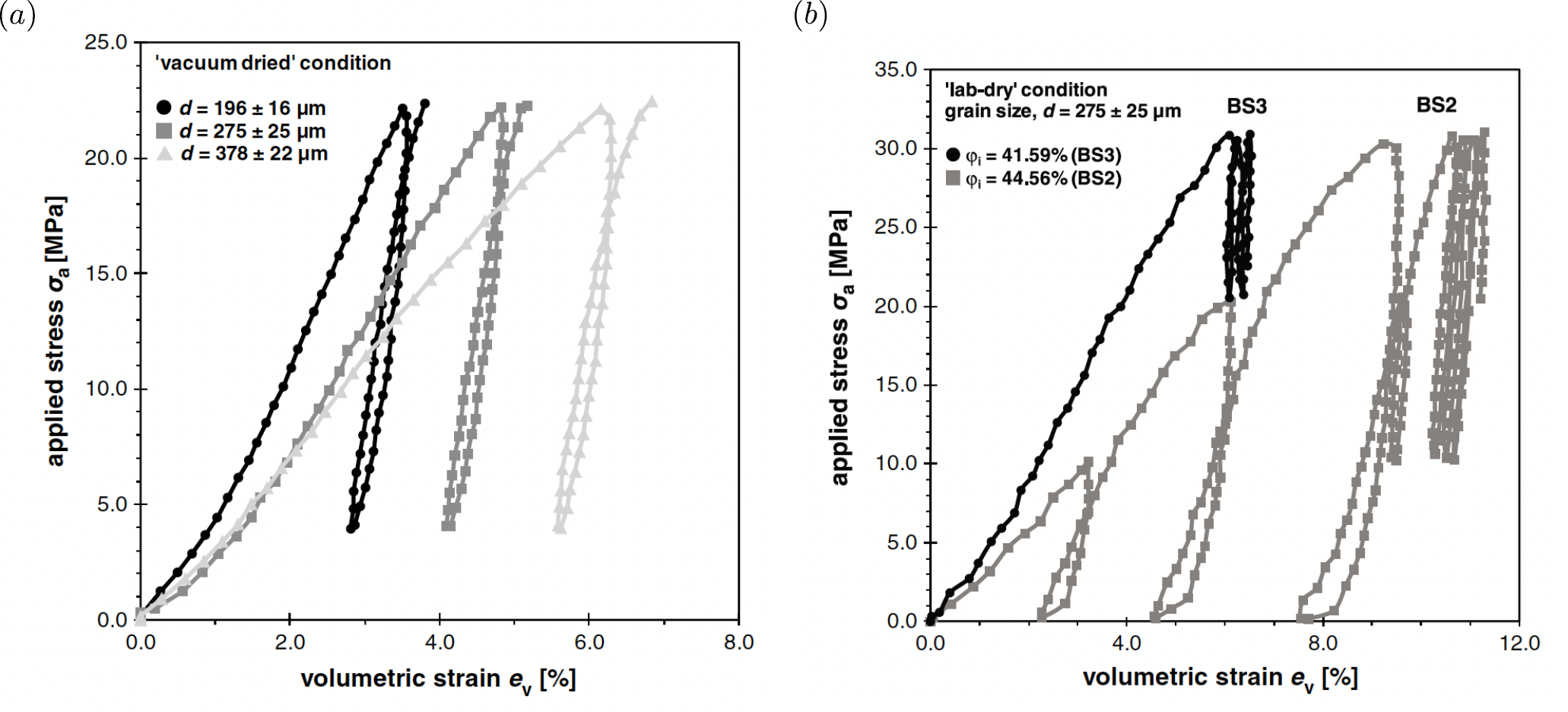}
    \caption{Experimental results on the effect of (a) grain diameter and (b) porosity from oedometer tests on sand, reproduced with permission from~\citet{Brzesowsky2014a}. Increasing porosity or grain diameter corresponds to less consolidated granular materials.}
    \label{fig::brzesowsky_oed_test}
\end{figure}

\begin{figure}[H]
    \centering
    \includegraphics[width=0.85\linewidth]{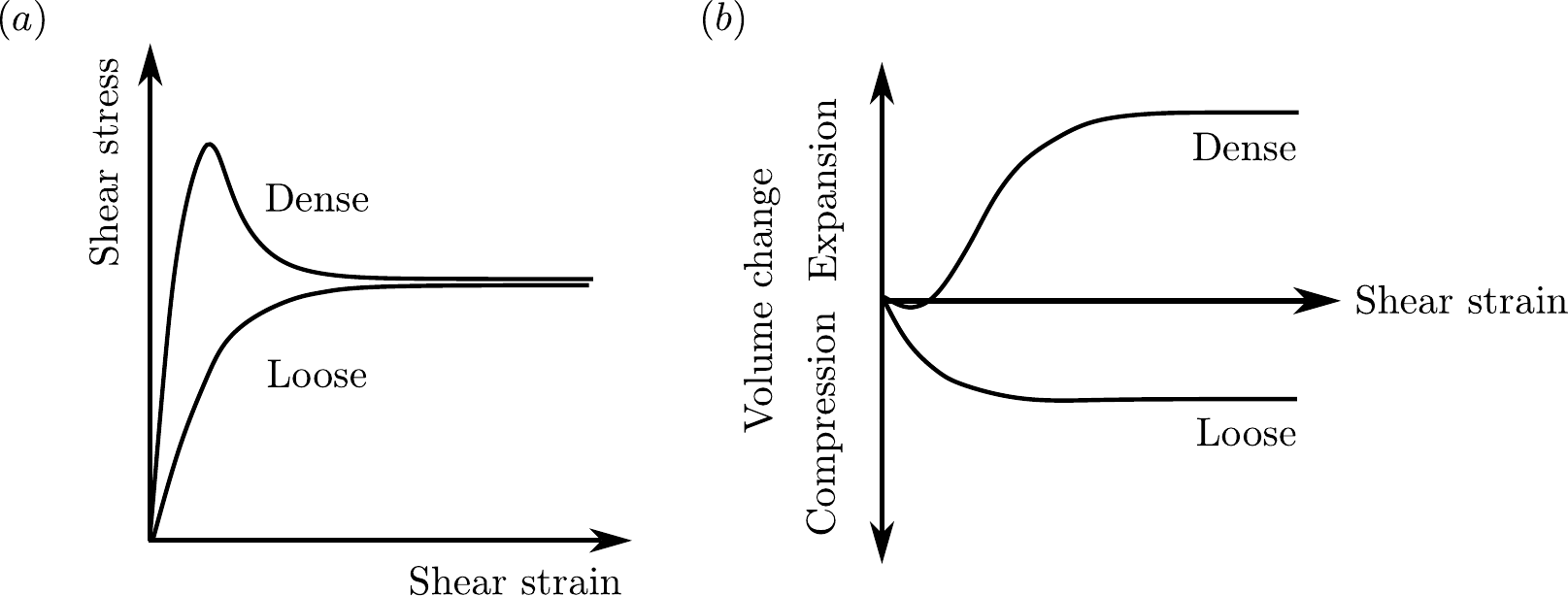}
    \caption{Schematic representation of the effect of consolidation on the (a) shear stress and (b) volumetric response of sheared granular layer. Dense granular assemblies dilate and soften, while loosely compacted layers harden and compact~\cite{Das2019}.}
    \label{fig::das_1983}
\end{figure}

\begin{figure}[H]
    \centering
    \includegraphics[width=0.85\linewidth]{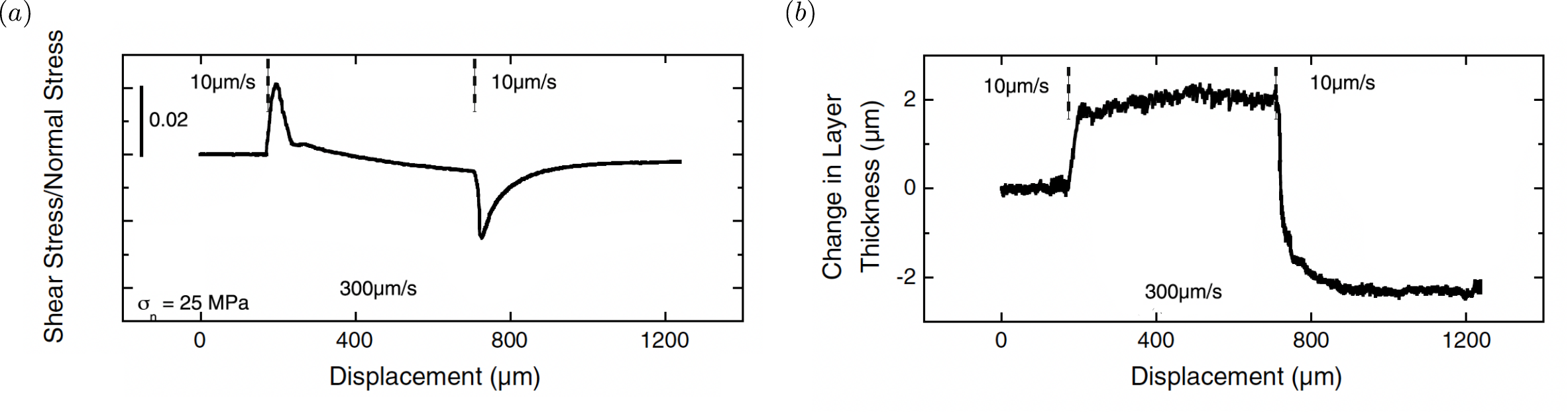}
    \caption{Response of a granular layer to stepped loading, adapted with permission from~\citet{Rathbun2013}. (a) The shear stress increases (decreases) with step increases (decreases) in the shear displacement rate. (b) The layer dilates (compacts) in response to faster (slower) loading rate. Both the shear stress and layer height approach a rate-dependent steady-state value at constant loading rate. }
    \label{fig::rathbun_marone}
\end{figure}

Granular materials exhibit a wealth of behaviors sensitive to pressure, consolidation, and rate. Here, we summarize the range of experimental observations we aim to capture in our critical-state model.
\begin{itemize}[itemsep=1mm, parsep=0pt]
    \item[(i)] More consolidated granular samples have a stiffer compressive response~\cite{Brzesowsky2014a}. This change in consolidation is achieved experimentally in~\citet{Brzesowsky2014a} through smaller grain diameters and decreased porosity of a granular layer (Figure~\ref{fig::brzesowsky_oed_test}).
    \item[(ii)] In response to shear, highly consolidated materials exhibit a dilatant, softening response while loosely consolidated materials compact and harden (Figure~\ref{fig::das_1983})~\cite{Das2019}.
    \item[(iii)] The shear resistance increases with the confining pressure~\cite{Byerlee1978}.
    \item[(iv)] In response to step changes in loading rate, the transient shear stress and volumetric responses depend on both strain and strain-rate (Figure~\ref{fig::rathbun_marone}(a))~\cite{Rathbun2013}.
    \item[(v)] The shear layer dilates (compacts) with increases (decreases) in loading rate (Figure~\ref{fig::rathbun_marone}(b)) \cite{Rathbun2013}.
    \item[(vi)] At long times at a constant loading rate, the stress and volume reach a rate-dependent steady-state value (Figure~\ref{fig::rathbun_marone})~\cite{Rathbun2013}.
\end{itemize}

\section{Governing equations}
\label{sec::governingEquations}

We present a general finite-kinematics framework for critical-state plasticity motivated by~\cite{ortiz2004variational}. Taking advantage of the variational structure of the governing equations, we identify effective stress-strain relations for a general rigid-plastic material. We then specialize the formulation to describe the shearing and dilating granular materials found in fault zones and their laboratory analogs.

\subsection*{Notational conventions}

We adopt the following notational conventions. We denote the $n\times n$ \textit{identity matrix} by $\bI$. We define the \textit{inner product} of two $n\times n$ matrices $\bA$ and $\bB$ as
\begin{equation}
    \bA \cdot \bB := \sum_{i=1}^n \sum_{j=1}^n A_{ij} B_{ij} .
\end{equation}
The \textit{trace} of an $n\times n$ matrix $\bA$ is then,
\begin{equation} 
    \tr\bA := \bA \cdot \bI = \sum_{i=1}^n A_{ii} ,
\end{equation}
and the \textit{norm} is 
\begin{equation} 
    |\bA| := \sqrt{\bA\cdot\bA} = \sqrt{\tr(\bA\bA^T)}.
\end{equation}
For any $n\times n$ matrix $\bA$ we use the notation
\begin{equation}
    \widebar{A} := \tr\bA,
    \quad
    \bA' := \bA - \frac{1}{3} \widebar{A} \, \bI ,
\end{equation}
to denote the \textit{volumetric} and \textit{deviatoric} parts of $\bA$, respectively. 

\subsection{Finite kinematics}
We assume a multiplicative decomposition of the deformation gradient,
\begin{equation}
    \bF = \Fe\Fp,
\end{equation}
into an elastic deformation $\Fe$ and a plastic deformation $\Fp$~\cite{Lee1964}.

\subsection{Rigid elasticity}
The yield stress of fault gouge in shear is pressure-dependent. At the depths at which earthquakes occur, the shear and normal stresses are on the order of $10-100$ MPa. As the elastic moduli of rock are on the order of $10$ GPa, the associated elastic deformation is very small (yield strains of $10^{-4}-10^{-3}$) compared to the finite strains achieved across a shear layer during an earthquake. Thus, we assume rigid elasticity, such that the right Cauchy-Green deformation tensor $\bC^e \equiv \bI$, or equivalently, the elastic deformation is a pure rotation, 
\begin{equation}
    \Fe = \Rel \in SO(3).
\end{equation}

\subsection{Volumetric plasticity}
We denote the Jacobian of the deformation by $J := \det \bF$. For rigid elasticity, $J^e \equiv 1$, and thus $J^p = J$. Then, we define the corresponding volumetric plastic strain by
\begin{equation} \label{eq::thp_Jp}
    \theta^p := \log J^p = \log J.
\end{equation}

\subsection{Rate of deformation}
The spatial velocity gradient $\bL = \dot{\bF}\bF^{-1}$ can be additively decomposed into
\begin{equation} \label{eq::LeLp}
    \bL = \Fedot\Fe^{-1} + \Fe (\Fpdot \Fp^{-1})\Fe^{-1} := \bL^e + \bL^p.
\end{equation}
For $\Fe \in SO(3)$, $\bL^e$ is skew-symmetric and, assuming no plastic spin, $\bL^p$ is symmetric,
\begin{equation} \label{6nU5BX}
    \bD^e := \operatorname{sym}\bL^e \equiv \bZ ,
    \quad
    \bW^p := \operatorname{skew}\bL^p \equiv \bZ .
\end{equation}
Then,
\begin{equation} \label{eq::L_decomp}
    \bW = \bW^e,\quad \bD = \bD^p.
\end{equation}

\subsection{Elastic deformation}
Due to the decomposition~\eqref{eq::L_decomp}, the elastic deformation is the solution of the ODE 
\begin{equation}
    \dot{\bR}^e \Rel^{-1} = \bW,
\end{equation}
with initial condition $\Rel(t = 0) = \bI$. Thus, $\Rel(t)$ is given by
\begin{equation} \label{eq::re_define}
    \Rel(t) = \exp\left(\int_0^t \bW(s) \ ds\right),
\end{equation}
where the matrix exponential of a skew-symmetric matrix is a rotation matrix.

\subsection{Flow rule}
The classical plastic-flow rules of small-strain rate-independent plasticity are derived from the elastic domain in stress space by invoking a normality rule~\cite{lubliner2008plasticity}. In finite deformations, this approach raises conflicts with requirements of material-frame indifference or objectivity. 

To sidestep these difficulties,~\citet{Ortiz1999} formulated a kinematical theory of finite-deformation plastic flow that subsumes the classical theory when restricted to small strains and relies on the principle of maximum dissipation to characterize yield, ensuring material-frame indifference. Thus, following~\cite{Ortiz1999}, we write the rate of plastic deformation as
\begin{equation} \label{eq::flow_rule}
    \Fpdot \Fp^{-1} 
    = 
    \dot{\eps}^p \, \bM,
    \quad
    \bM^T = \bM ,
\end{equation}
where $\dot{\eps}^p$ is the plastic strain rate and $\bM$ is a symmetric kinematic tensor that sets the direction of plastic flow. Taking derivatives in~\eqref{eq::thp_Jp}, the volumetric plastic strain rate is
\begin{equation} \label{eq::flow_thp}
    \dot{\theta}^p 
    = 
    \tr\left(\Fpdot \Fp^{-1}\right) 
    = 
    \dot{\eps}^p \, \widebar{M}, 
\end{equation}
whereupon~\eqref{eq::flow_rule} can be rewritten as
\begin{equation} \label{DNZaNe}
    \Fpdot \Fp^{-1} 
    = 
    \dfrac{1}{3} \, \dot{\theta}^p \, \bI
    +
    \dot{\eps}^p \, \bM' .
\end{equation}

Specific classes of finite-deformation plastic materials can be conveniently modeled by appending suitable constraints to the tensor $\bM$, effectively setting forth particular forms of the plastic flow rule~\cite{Ortiz1999}. The most general plastic-flow constraint can be expressed as
\begin{equation} \label{HXRFef}
    g(\bM) = 0 ,
\end{equation}
where $g$ is a scalar-valued function. Examples of choices of the constraint (\ref{HXRFef}) that return classical plasticity models are given in~\cite{Ortiz1999}.

\subsection{Maximum dissipation}
The rate and direction of plastic flow are posited to maximize dissipation, in accordance with the principle of maximum dissipation~\cite{lubliner2008plasticity}. To this end, one formulates a dissipation function that depends on the kinematic tensor $\bM$ and plastic strain rate $\dot{\eps}^p$, such that minimizing the function with respect to both arguments maximizes the dissipation.

Here, we notice that the plastic-flow constraint,~\eqref{HXRFef}, sets forth an implicit dependence of $\dot{\eps}^p$ on $\bD$ due to the assumption of rigid elasticity. That is, inserting~\eqref{eq::flow_rule} into~\eqref{HXRFef} and using~\eqref{eq::L_decomp} gives
\begin{equation} \label{sPkhqE}
    g(\Rel^{-1}(\bD/\dot{\eps}^p) \, \Rel) = 0 ,
\end{equation}
which implicitly defines a function $\dot{\eps}^p(\bD)$, as surmised. Hence, the dissipation function can be written entirely in terms of $\bD$ instead of $\bM$ and $\dot{\eps}^p$. In addition, differentiating \eqref{sPkhqE} with respect to $\bD$, we obtain
\begin{equation} 
    \frac{1}{\dot{\eps}^p} \Rel
    \frac{\partial g}{\partial\bM} \, \Rel^{-1}
    -
    \frac{1}{\dot{\eps}^{p \, 2}}
    \Big(\Rel
    \frac{\partial g}{\partial\bM} \, \Rel^{-1}\cdot\bD\Big)
    \frac{\partial\dot{\eps}^p}{\partial\bD}
    = 0 ,
\end{equation}
whence, solving, the derivative of $\dot{\eps}^p(\bD)$ follows as
\begin{equation} \label{07HNrr}
    \frac{\partial\dot{\eps}^p}{\partial\bD}
    =  
    \dot{\eps}^p
    \Big(\Rel
    \frac{\partial g}{\partial\bM} \, \Rel^{-1}\cdot\bD\Big)^{-1}
    \Big(\Rel
    \frac{\partial g}{\partial\bM} \,\Rel^{-1}\Big).
\end{equation}

In formulating the dissipation function, we assume that the stored energy $W^p$ is solely a function of the plastic volumetric deformation $\theta^p$ and temperature $T$ and define the consolidation pressure $p_0(\theta^p, T)$ by
\begin{equation}
    p_0(\theta^p, T) 
    :=
    \dfrac{\partial W^p}{\partial\theta^p}(\theta^p, T).
\end{equation} 
This assumption is motivated by theories of free volume in granular media and amorphous metals \cite{cohen1959molecular,edwards1989theory} and is born out by the validation examples in Section~\ref{sec::results}. Likewise, we assume a kinetic potential of the form ${\psi}(\dot{\theta}^p, \dot{\eps}^p;\theta^p, T)$, depending on the state only through $(\theta^p,T)$. 

Building on these relations, we postulate a dissipation function of the general form
\begin{equation} \label{eq::max_diss2}
    f(\bD;\sig,\Rel;\theta^p,T) 
    = 
    -
    \sig\cdot \bD 
    + 
    p_0(\theta^p, T) \, \dot{\theta}^p
    + 
    {\psi}(\dot{\theta}^p, \dot{\eps}^p;\theta^p, T),
\end{equation}
for fixed Cauchy stress tensor $\sig$, elastic rotation $\Rel$, volumetric plastic strain $\theta^p$ and temperature $T$. In~\eqref{eq::max_diss2}, $\dot{\eps}^p$ and $\dot{\theta}^p$ are understood to be functions of $\bD$ through~\eqref{eq::flow_thp} and~\eqref{sPkhqE}. As a result, both the rate and direction of plastic flow follow from minimizing~\eqref{eq::max_diss2} with respect to $\bD$. The function $p_0(\theta^p, T)$ represents a critical or yield pressure for plastic volumetric deformation and ${\psi}(\dot{\theta}^p, \dot{\eps}^p;\theta^p, T)$ is a dissipation potential, with both to be characterized for specific classes of materials. 

For simplicity of notation, we subsequently omit any and all dependencies on the state variables $(\theta^p, T)$ and leave them implicit throughout. 

Minimizing $f$ with respect to $\bD$ gives the effective stress-strain relation
\begin{equation} \label{7uqQeL}
    \sig(\bD)
    = 
    \Big(
        p_0 
        + 
        \dfrac{\partial {\psi}}{\partial \dot{\theta}^p}(\widebar{D}, \dot{\eps}^p(\bD)) 
    \Big) \,
    \bI 
    + 
    \dfrac{\partial {\psi}}{\partial \dot{\eps}^p}(\widebar{D}, \dot{\eps}^p(\bD))
    \dfrac{\partial \dot{\eps}^p}{\partial \bD},
\end{equation}
where $\dot{\eps}^p(\bD)$ is defined implicitly by~\eqref{sPkhqE} and its derivative is given by~\eqref{07HNrr}. As expected, the stress-strain relations~\eqref{7uqQeL} have a variational structure 
\begin{equation} \label{Krh0pX}
    \sig(\bD) = \frac{\partial \Psi}{\partial\bD}(\bD) ,
\end{equation}
with effective dissipation potential
\begin{equation}
    \Psi(\bD)
    =
    p_0 \widebar{D}
    +
    \psi(\widebar{D}, \dot{\eps}^p(\bD)) .
\end{equation}
If the dissipation potential ${\psi}(\dot{\theta}^p, \dot{\eps}^p)$ is convex, then so is the
effective dissipation potential $\Psi(\bD)$. However, $\Psi(\bD)$ may fail to be differentiable at $\bD = \mathbf{0}$, in which case the derivative \eqref{Krh0pX} is valid for $\bD \neq \mathbf{0}$ only and must be carefully reinterpreted in the sense of subdifferentials~\cite{rockafellar2015convex}, which may be set-valued, at $\bD = \mathbf{0}$. We recall that the subdifferential of $\Psi(\bD)$ at $\bD = \mathbf{0}$ is the set in stress space
\begin{equation} \label{q4gHIv}
    \partial\Psi(\mathbf{0})
    :=
    \{
        \sig \, : \, \Psi(\bD) \geq \sig \cdot \bD ,
        \;\, \text{for all} \; \bD
    \} ,
\end{equation}
where we assume that $\psi(\dot{\theta}^p, \dot{\eps}^p)$ is convex and differentiable away from $(0,0)$ and $\psi(0,0) = 0$. Therefore, we identify the elastic domain with the set~\eqref{q4gHIv} in stress space. 

We note that the elastic domain is generally dependent on the state of volumetric plastic deformation $\theta^p$ due to consolidation and on the elastic rotation $\Rel$ due to anisotropy. 

\subsection{Specialization of plastic flow potential to critical state models}

Further explicit representations can be derived in special cases. For instance, if the plastic-flow rule is isotropic and independent of the third invariant of $\bM$, then the general plastic-flow constraint~\eqref{HXRFef} specializes to
\begin{equation} \label{eq::flow_potential}
    g(\widebar{M}, |\bM'|) = 0 . 
\end{equation}

Under this assumption, we have the identity
\begin{equation}
    \dfrac{\partial g}{\partial \bM}
    =
    \dfrac{\partial g}{\partial \widebar{M}}
    \, \bI
    +
    \dfrac{\partial g}{\partial |\bM'|}
    \dfrac{\bM'}{|\bM'|} ,
\end{equation}
which can be used, e.~g., in~\eqref{07HNrr}.

\begin{figure}[h]
    \centering
    \includegraphics[width=0.4\linewidth]{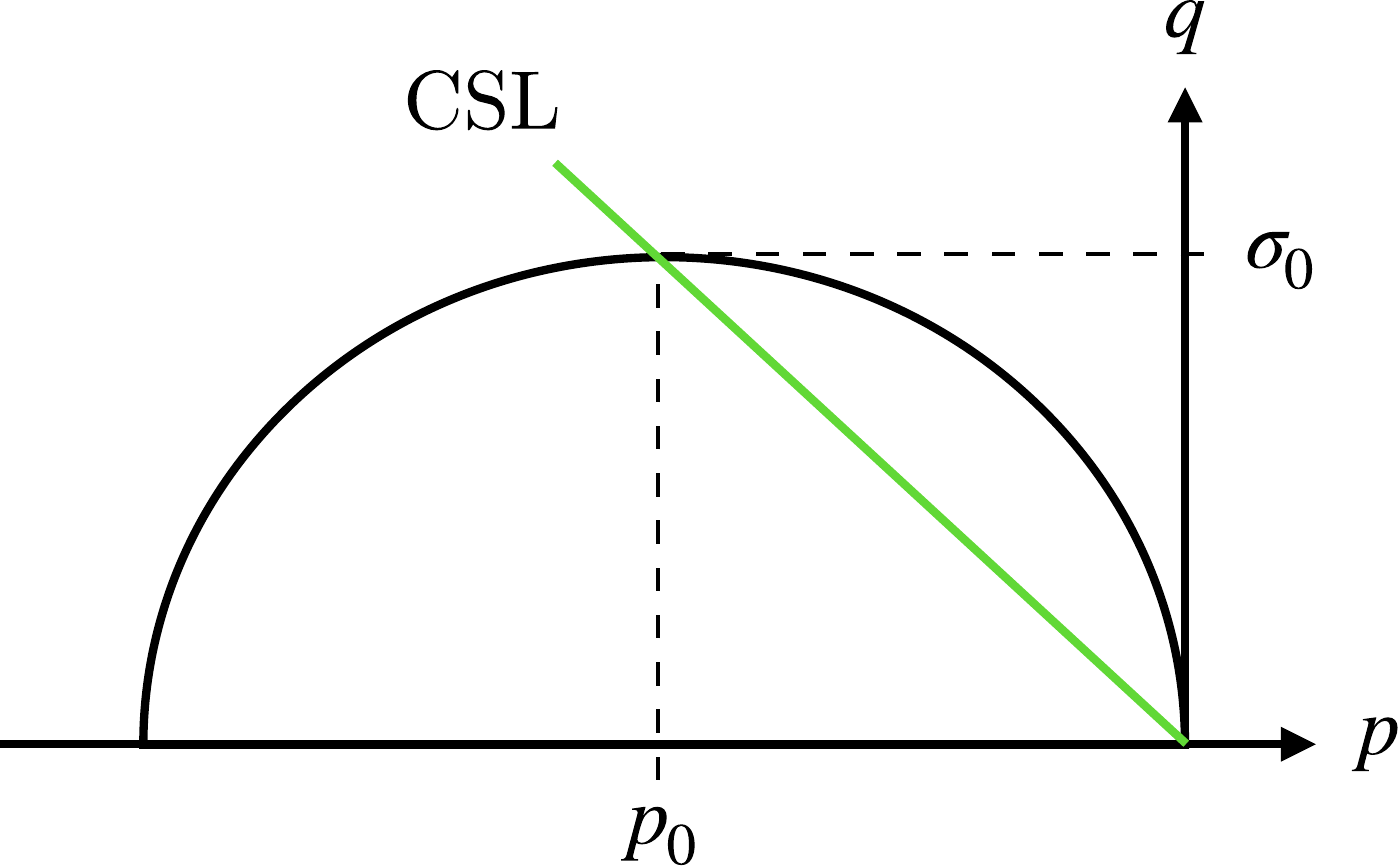}
    \caption{Elliptic yield surface in the $(p,q)$ plane. The critical state line (CSL) corresponds to the stress states for which the material response is isochoric.}
    \label{fig::yield_surf}
\end{figure}

A further particular example is the Cam-Clay model, which can be characterized by a plastic flow constraint of the form~\cite{ortiz2004variational} 
\begin{equation} \label{eq::cc_kinetic_pot}
    g(\widebar{M}, |\bM'|) 
    = 
    \dfrac{\widebar{M}^2}{\alpha^2} + \dfrac{2}{3}|\bM'|^2 - 1 = 0,
\end{equation}
where $\alpha>0$ is an internal friction coefficient related to the friction angle of the Mohr-Coulomb failure criteria~\cite{ortiz2004variational}. In this case, the implicit relation~\eqref{sPkhqE} gives the explicit relation
\begin{equation} \label{eq::ep_D}
    \dot{\eps}^p(\bD)
    = 
    \Bigg(
        \dfrac{\widebar{D}^2}{\alpha^2} + \dfrac{2}{3}|\bD'|^2
    \Bigg)^{1/2} ,
\end{equation}
with derivative
\begin{equation}
    \dfrac{\partial \dot{\eps}^p}{\partial \bD} 
    = 
    \frac{1}{\dot{\eps}^p(\bD)}
    \Bigg(
        \dfrac{\widebar{D}}{\alpha^2}\bI + \dfrac{2}{3} \bD'
    \Bigg) ,
\end{equation}
and the effective stress-strain relation~\eqref{7uqQeL} takes the form
\begin{equation} \label{eq::stress_strain}
    \sig(\bD) 
    = 
    p \, \bI 
    + 
    \dfrac{2}{3}
    \dfrac{\bD'}{\dot{\eps}^p(\bD)} 
    \dfrac{\partial {\psi}}{\partial \dot{\eps}^p}(\widebar{D}, \dot{\eps}^p(\bD)),
\end{equation}
where
\begin{equation} \label{eq::p_D}
    p = p_0 + \dfrac{\partial {\psi}}{\partial \dot{\theta}^p}(\widebar{D}, \dot{\eps}^p(\bD)) + \dfrac{\widebar{D}}{\alpha^2 \dot{\eps}^p(\bD)}\dfrac{\partial {\psi}}{\partial \dot{\eps}^p}(\widebar{D}, \dot{\eps}^p(\bD)).
\end{equation}
Taking norms of the deviatoric part of~\eqref{eq::stress_strain} gives the scalar relation
\begin{equation} \label{eq::q_D}
    q = \sqrt{\frac{2}{3}}
    \dfrac{|\bD'|}{\dot{\eps}^p(\bD)} 
    \dfrac{\partial {\psi}}{\partial \dot{\eps}^p}(\widebar{D}, \dot{\eps}^p(\bD)), 
    \quad 
    q := \sqrt{\frac{3}{2}} | \sig' |,
\end{equation}
where $q$ is the Mises stress. Using the stress invariants~\eqref{eq::p_D} and~\eqref{eq::q_D},~\eqref{eq::ep_D} becomes
\begin{equation}
    \alpha^2 \Big(p- p_0 - \dfrac{\partial {\psi}}{\partial \dot{\theta}^p}(\widebar{D}, \dot{\eps}^p(\bD))\Big)^2 
    + 
    q^2 
    = 
    \Big(\dfrac{\partial {\psi}}{\partial \dot{\eps}^p}(\widebar{D}, \dot{\eps}^p(\bD))\Big)^2.
\end{equation}
When $\bD = \bZ$, $\dot{\eps}^p(\bZ) = 0$ and the elastic domain reduces to
\begin{equation} \label{eq::rel_yield}
    \sqrt{\alpha^2(p-p_0)^2 + q^2}-\sigma_0 \leq 0,
\end{equation}
where the yield surface corresponds to the stress states for which the equality is achieved, and we assume
\begin{equation} \label{eq::pdv_psi}
    \dfrac{\partial {\psi}}{\partial \dot{\theta}^p} (0,0) = 0 ,
    \quad
    \dfrac{\partial {\psi}}{\partial \dot{\eps}^p} (0,0) = \sigma_0 .
\end{equation}
In this case, the yield surface~\eqref{eq::rel_yield} is an ellipse and the critical state line is a straight line (Figure~\ref{fig::yield_surf}).

\section{Specialization to a granular layer}
\label{sec::shear_layer}

\begin{figure}[t]
    \centering
    \includegraphics[width=0.85\linewidth]{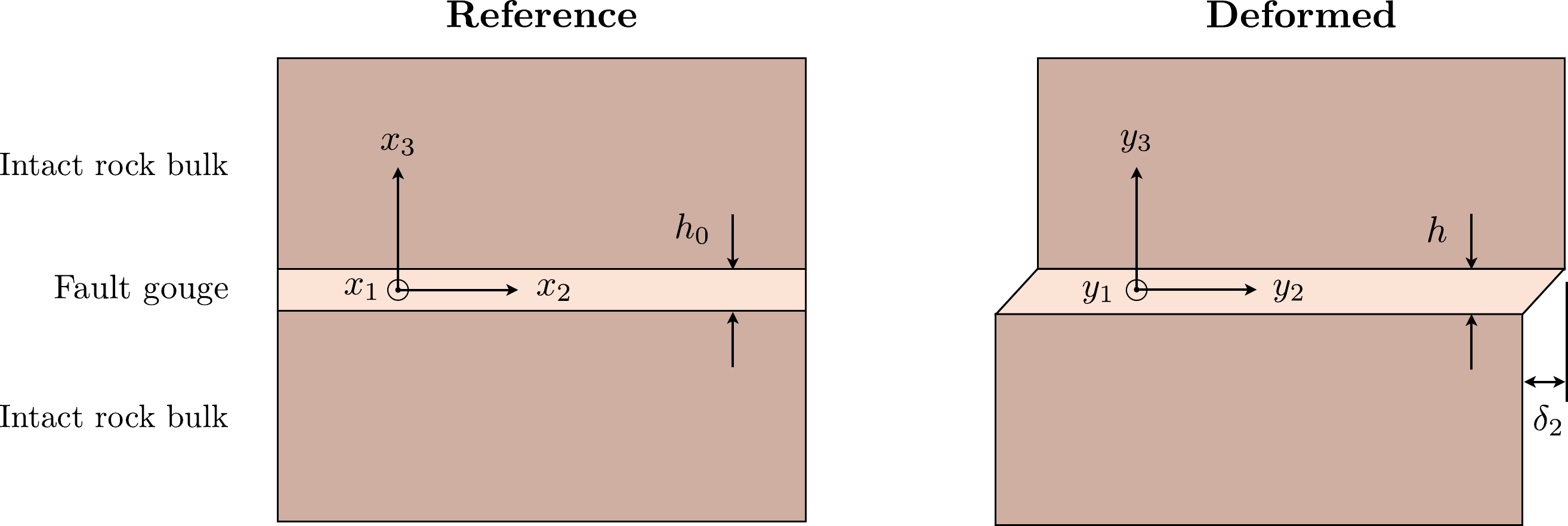}
    \caption{Shearing and dilating shear layer of fault gouge with initial height $h_0$ and deformed height $h$.}
    \label{fig:shearlayer_schem}
\end{figure}

To apply the general framework to the granular material found in fault zones, we consider shear and dilation of a granular layer of fault gouge of initial thickness $h_0$, shown in Figure~\ref{fig:shearlayer_schem}. The deformation mapping within the layer has the form
\begin{equation}
    y_i(\mathbf{x}) = x_i + \dfrac{\delta_i}{h_0}x_3,
\end{equation}
where $i = 1,2,3$. Then, the jump in displacement across the layer is
\begin{equation}
    \delta_i = y_i(\mathbf{x}_p+(h_0/2)\be_3) - y_i(\mathbf{x}_p-(h_0/2)\be_3),
\end{equation}
where $\mathbf{x}_p = (x_1, x_2, 0)$ is the in-plane position vector. Similarly, the thickness of the layer at time $t$ is
\begin{equation} \label{eq::h(t)}
    h := h_0 + \delta_3.
\end{equation}
The deformation gradient follows as
\begin{equation}
    F_{ij} = \delta_{ij} + \frac{\delta_i}{h_0} \delta_{3j} ,
    \quad
    \bF = \bI + \frac{\del}{h_0} \otimes \be_3 ,
    \quad
    \bF = \begin{pmatrix}
        1 & 0 & \delta_1/h_0 \\
        0 & 1 & \delta_2/h_0 \\
        0 & 0 & 1 + \delta_3/h_0
    \end{pmatrix} ,
\end{equation}
in indicial, invariant, and matrix notation, respectively, and has determinant 
\begin{equation}
    J = \det \bF = 1 + \dfrac{\delta_3}{h_0} = \dfrac{h}{h_0}.
\end{equation}
From~\eqref{eq::thp_Jp}, the layer dilation is directly related to the volumetric plastic strain through
\begin{equation} \label{eq::h_thp}
    \theta^p = \log J  = \log \dfrac{h}{h_0} ,
    \quad
    h = h_0 \, \exp{\theta^p} .
\end{equation}

For the deformation considered, the spatial velocity gradient $\bL$ is
\begin{equation}
    L_{ij} = \frac{\dot{\delta}_i}{h} \delta_{3j} ,
    \quad 
    \bL = \frac{\dot{\del}}{h} \otimes \be_3 ,
    \quad
    \bL = \begin{pmatrix}
        0 & 0 & \dot{\delta}_1/h  \\ 0 & 0 & \dot{\delta}_2/h \\ 0  & 0  & \dot{\delta}_3/h 
    \end{pmatrix},
\end{equation}
and the skew-symmetric and symmetric parts are given, respectively, by the spin tensor,
\begin{equation}
    W_{ij}
    =
    \frac{1}{2} 
    \left(
        \frac{\dot{\delta}_i}{h} \delta_{3j}
        -
        \frac{\dot{\delta}_j}{h} \delta_{3i}
    \right) ,
    \;\;
    \bW = 
    \frac{1}{2}
    \left(
        \frac{\dot{\del}}{h} \otimes \be_3 
        -
        \be_3 \otimes \frac{\dot{\del}}{h}   
    \right) ,
    \;\;
    \bW 
    = 
    \begin{pmatrix}
        0 & 0 & {\dot{\delta}_1}/{2h}  \\ 
        0 & 0 & {\dot{\delta}_2}/{2h} \\ 
        -{\dot{\delta}_1}/{2h}  & -{\dot{\delta}_2}/{2h} & 0
    \end{pmatrix} ,
\end{equation}
and the rate-of-deformation tensor,
\begin{equation} \label{eq::DLp_gen}
    D_{ij}
    =
    \frac{1}{2} 
    \left(
        \frac{\dot{\delta}_i}{h} \delta_{3j}
        +
        \frac{\dot{\delta}_j}{h} \delta_{3i}
    \right) ,
    \;\;
    \bD = 
    \frac{1}{2}
    \left(
        \frac{\dot{\del}}{h} \otimes \be_3 
        +
        \be_3 \otimes \frac{\dot{\del}}{h}   
    \right) ,
    \;\;
    \bD 
    = 
    \begin{pmatrix}
        0 & 0 & {\dot{\delta}_1}/{2h}  \\ 
        0 & 0 & {\dot{\delta}_2}/{2h}  \\ 
        {\dot{\delta}_1}/{2h} & {\dot{\delta}_2}/{2h} & {\dot{\delta}_3}/{h}
    \end{pmatrix} .
\end{equation}
At fixed $\Fe$, the stress tensor finally follows by inserting \eqref{eq::DLp_gen} into \eqref{7uqQeL}. We note that, in general, the in-plane components of the stress tensor are not zero due to the active pressure that results from normal stress. 

For a shearing and dilating granular layer, $\bD$ is fully parameterized by the deformation rates 
\begin{equation}
    \dot{\gamma}_i := \frac{\dot{\delta}_i}{h} .
\end{equation}
Hence, the stress-strain relations \eqref{7uqQeL} reduce to functions of the form
\begin{equation} 
    \sig = \sig(\dot{\gam}) = \sig(\dot{\del}/h),
\end{equation}
where the state dependence is implicit. In particular, the tractions 
\begin{equation}
    \boldsymbol{\tau} := \boldsymbol{\sigma}\,\be_3 ,
\end{equation}
which are work-conjugate to $\dot{\gam}$, obey a law of the form
\begin{equation} \label{7MA1Ed}
    \boldsymbol{\tau} 
    = 
    \sig(\dot{\gam}) \, \be_3 
    := 
    \boldsymbol{\tau}(\dot{\gam}).
\end{equation}
The work-conjugacy between $\boldsymbol{\tau}$ and $\dot{\gam}$ has the immediate consequence that the stress-strain law~\eqref{7MA1Ed} derives from a potential, namely,
\begin{equation}
    \boldsymbol{\tau}(\dot{\gam})
    =
    \frac{\partial\Phi}{\partial\dot{\gam}}(\dot{\gam}) ,
    \quad
    \Phi(\dot{\gam}) = \Psi(\bD) ,
\end{equation}
with $\bD$ as in~\eqref{eq::DLp_gen}. 

When regarded as a relation for $\boldsymbol{\tau}(\dot{\del})$ at given $h$, with the dependence on state implicit,~\eqref{7MA1Ed} expresses a stress-strain law of the type commonly assumed in rate-and-state models of fault shear resistance~\cite{Rabinowicz1951, Rabinowicz1958, Dieterich1978, Ruina1983}. However, it bears emphasis that the shear resistance law thus obtained is not postulated \textit{ad hoc}, but rather derived as the effective behavior of a shearing and dilating layer composed of a granular material. 

\subsection{Specialization to Cam-Clay}
We apply the Cam-Clay plasticity model to the particular deformation of a shearing and dilating granular layer discussed above. Using~\eqref{eq::DLp_gen} in~\eqref{eq::ep_D},
\begin{equation}\label{eq::ep_shearlayer}
    \dot{\varepsilon}^p(\dot{\gam}) = \sqrt{\frac{\dot{\gamma}_1^2}{3} + \frac{\dot{\gamma}_2^2}{3} + \frac{\dot{\gamma}_3^2}{\tilde{\alpha}^2}},
\end{equation}
where we write
\begin{equation}
    \tilde{\alpha} := \frac{\alpha}{\sqrt{1+\frac{4}{9}\alpha^2}},
\end{equation}
and, from~\eqref{eq::flow_thp}, $\dot{\theta}^p = \widebar{D} = \dot{\gamma}_3$. From~\eqref{eq::DLp_gen}, the corresponding stress-strain relation~\eqref{eq::stress_strain} becomes
\begin{equation}
\begin{aligned}
    \sig(\dot{\gam}) &= 
    \left(p_0 
    + 
    \frac{\partial \psi}{\partial \dot{\theta}^p}(\dot{\gamma}_3, \dot{\eps}^p(\dot{\gam})) 
    + 
    \frac{\dot{\gamma}_3}{\dot{\eps}^p(\dot{\gam})}\left(\frac{1}{\tilde{\alpha}^2}
    - 
    \dfrac{2}{3}\right)\frac{\partial {\psi}}{\partial \dot{\eps}^p}(\dot{\gamma}_3, \dot{\eps}^p(\dot{\gam}))\right) \mathbf{I} \\
    &+ \frac{1}{3\dot{\eps}^p(\dot{\gam})}(\dot{\gam} \otimes \be_3 
    + 
    \be_3 \otimes \dot{\gam})\frac{\partial {\psi}}{\partial \dot{\eps}^p}(\dot{\gamma}_3, \dot{\eps}^p(\dot{\gam})).
\label{eq::sig_shearlayer}
\end{aligned}
\end{equation}
As a result, the tractions $\boldsymbol{\tau}$ are given by
\begin{equation} \label{eq::tau_shearlayer}
    \boldsymbol{\tau}(\dot{\gam}) = 
    \left(p_0 
    + 
    \frac{\partial \psi}{\partial \dot{\theta}^p} (\dot{\gamma}_3, \dot{\eps}^p(\dot{\gam}))\right)\be_3 
    + 
    \frac{1}{\dot{\eps}^p(\dot{\gam})}
    \left(
        \frac{\dot{\gamma}_3}{\tilde{\alpha}^2}\be_3 
        + 
        \frac{\dot{\gam} - \dot{\gamma}_3\be_3}{3}
    \right)
    \frac{\partial \psi}{\partial \dot{\eps}^p} (\dot{\gamma}_3, \dot{\eps}^p(\dot{\gam})),
\end{equation}
and, by symmetry, the lateral stresses are
\begin{equation}
    \sigma_{\alpha \beta} (\dot{\gam})
    = 
    \Big(
        p_0 
        + 
        \frac{\partial \psi}{\partial \dot{\theta}^p}(\dot{\gamma}_3, \dot{\eps}^p(\dot{\gam})) 
        + 
        \frac{\dot{\gamma}_3}{\dot{\eps}^p(\dot{\gam})}\left(\frac{1}{\tilde{\alpha}^2} 
        - 
        \dfrac{2}{3}\right)\frac{\partial {\psi}}{\partial \dot{\eps}^p}(\dot{\gamma}_3, \dot{\eps}^p(\dot{\gam}))
    \Big)
    \, \delta_{\alpha\beta} .
\end{equation}
It follows that the confining in-plane stresses are independent of direction, as expected from isotropy.

A number of further restrictions on the layer behavior can be gleaned from standardized experiments on soils and sands. In some cases, such restrictions suffice to fully identify the behavior of the layer, as shown next.

\subsection{Oedometer test}
\label{sec::restricted}
Oedometer tests relate the consolidation stress of a granular material to its volume change. Here, we use these experiments to restrict the functional forms of the consolidation pressure $p_0$ and rate potential $\psi(\dot{\theta}^p, \dot{\eps}^p)$.

In an oedometer test, the granular material is laterally confined such that $\dot{\gamma}_1 = \dot{\gamma}_2 \equiv 0$. The test may be displacement controlled ($\dot{\gamma}_3<0$ prescribed) or load controlled ($\tau_{3} < 0 $ prescribed). Consider the displacement-control case. For this deformation,~\eqref{eq::ep_shearlayer} has the simplified form
\begin{equation}
    \dot{\eps}^p(\dot{\gamma}_3) = \dfrac{|\dot{\gamma}_3|}{\tilde{\alpha}},
\end{equation}
for given $\dot{\gamma}_3$, and the only nonzero component of the traction~\eqref{eq::tau_shearlayer} is
\begin{equation} \label{eq::oed_stress}
    \tau_{3}(\dot{\gamma}_3) 
    := 
    \boldsymbol{\tau}(\dot{\gamma}_3\be_3) \cdot \be_3 
    = 
    p_0 
    + 
    \dfrac{\partial {\psi}}{\partial \dot{\theta}^p}(\dot{\gamma}_3, \dot{\eps}^p(\dot{\gamma}_3)) 
    + 
    \dfrac{1}{\tilde{\alpha}} \operatorname{sgn}(\dot{\gamma}_3)
    \dfrac{\partial {\psi}}{\partial \dot{\eps}^p}(\dot{\gamma}_3, \dot{\eps}^p(\dot{\gamma}_3)).
\end{equation}
If we further assume that in the limit $\dot{\eps}^p \to 0$, $\tau_{3} = 0$ for $\dot{\gamma}_3>0$ (opening), i.~e., cohesive effects are solely viscous, then~\eqref{eq::oed_stress} and~\eqref{eq::pdv_psi} give that the yield stress $\sigma_0$ is related to the consolidation pressure $p_0$ through
\begin{equation} \label{eq::oed_restrict}
    \sigma_0 = - \tilde{\alpha} p_0.
\end{equation}

Experiments on granular media find a linear relation between the normal strain, $\delta_3/h_0$, and the logarithm of the normal stress, $\log (-\tau_3)$, at constant temperature and loading rate~\cite{Bjerrum1967, Leroueil1985, Karig2003, Choens2018}. The consolidation curve shifts with changes in temperature and loading rate~\cite{Suklje1957, Leroueil1985,Laloui2008}. For a fixed vertical strain, the consolidation stress increases at higher strain rates and lower temperatures; i.~e., larger normal stresses are required to achieve the same void ratio in the granular material. Additionally,~\citet{Leroueil1985} empirically found that, at small strain rates and large normal stresses relative to the pre-consolidation pressure, the normal stress is a separable function of the normal strain and the normal strain rate.

To match these experiments, we assume that $p_0$ and ${\psi}$ have the same $\theta^p$ and $T$ dependence, and choose the functional forms,
\begin{equation}
\begin{aligned}
    &
    p_0(\theta^p, T) 
    = 
    \dfrac{p_c(T)}{2} \exp\left(-\dfrac{\theta^p}{\theta^p_{\rm ref}}\right), 
    \\ &
    {\psi}(\dot{\theta}^p,\dot{\eps}^p; \theta^p, T) 
    = 
    \sigma_0 \, \dot{\eps}^p 
    + 
    p_0(\theta^p, T) \, 
   \tilde{\varphi}(\dot{\eps}^p)\, \dot{\theta}^p 
    + 
    \sigma_c(T)  \,
    \varphi(\dot{\eps}^p),
\end{aligned}
\label{eq::p0_exp}
\end{equation}
where $p_c(T)<0$ is the quasi-static pre-consolidation pressure with $\sigma_c := - \tilde{\alpha}p_c > 0$, 
$\theta^p_{\rm ref}>0$ is the reference volumetric strain, and $\tilde{\varphi}$ and $\varphi$ are rate potentials that are constrained by the convexity of $\psi(\cdot, \cdot)$~(\ref{sec::app_convexity}) and have to be selected for specific materials. 

As a result, inserting the representation~\eqref{eq::p0_exp} into~\eqref{eq::oed_stress},
\begin{equation}
    \tau_{3}(\dot{\gamma}_3) = p_0 + p_0 (\tilde{\varphi}(\dot{\eps}^p) + \dot{\eps}^p \tilde{\varphi}'(\dot{\eps}^p)) - \operatorname{sgn}(\dot{\gamma}_3)\left(p_0 + p_c \varphi'(\dot{\eps}^p)\right),
\end{equation}
such that in tension, $\operatorname{sgn}(\dot{\gamma}_3) = +1$,
\begin{equation}
    \tau_{3}(\dot{\gamma}_3>0) = p_0 (\tilde{\varphi}(\dot{\eps}^p) + \dot{\eps}^p\tilde{\varphi}'(\dot{\eps}^p)) - p_c\varphi'(\dot{\eps}^p),
\end{equation}
and in compression, $\operatorname{sgn}(\dot{\gamma}_3) = -1$,
\begin{equation}
    \tau_{3}(\dot{\gamma}_3<0) = p_0 (2 + \tilde{\varphi}(\dot{\eps}^p) + \dot{\eps}^p\tilde{\varphi}'(\dot{\eps}^p)) + p_c\varphi'(\dot{\eps}^p).
\label{eq::consol_test}
\end{equation}
Due to the exponential dependence of $p_0$ on $\theta^p$, there is a linear relation between the vertical strain and the logarithm of the vertical stress, and the effective consolidation pressure is rate-dependent, as desired.

The stress-strain relation~\eqref{eq::tau_shearlayer} is fully-calibrated solely considering experimental observations of oedometer tests, but must be solved numerically for a general deformation. In the following section, we make simplifying assumptions motivated by experiments that allow us to develop closed form expressions for $\boldsymbol{\tau}(\dot{\gam})$ for a shearing granular layer.

\subsection{Shear test}
\label{sec::shear_test}
We examine the response of the material model to a plane-strain ($\delta_1 = 0$), simple shear deformation, with prescribed shearing rate $\dot{\delta}_2>0$ and constant normal stress $-\tau_{3}:=\sigma_n>0$, with the goal of deriving closed-form expressions for the shear stress and state variable. The condition of constant normal stress is typical in experiments on the shear resistance of gouge layers~\cite{Dieterich1979, Ruina1983, marone1990frictional,Bedford2021} to mimic conditions of confinement by overburden on natural faults.

The plastic strain rate in simple shear depends on the shear loading rate and layer dilation rate through
\begin{equation}
    \dot{\eps}^p(\dot{\gamma}_2, \dot{\gamma}_3) = \sqrt{\dfrac{\dot{\gamma}_2^2}{3} + \dfrac{\dot{\gamma}_3^2}{\tilde{\alpha}^2}},
\label{eq::def_epdots}
\end{equation}
where $\dot{\eps}^p$ depends implicitly on $\theta^p$ due to the Eulerian nature of $\dot{\gam}$. 

Under plane-strain conditions, the shear tractions from~\eqref{eq::tau_shearlayer} are $\tau_1 = 0$ and
\begin{equation}
    \tau_2 = \dfrac{1}{\sqrt{3}} \dfrac{\partial \psi}{\partial \dot{\eps}^p}(\dot{\theta}^p, \dot{\eps}^p),
\label{eq::tau2_gen}
\end{equation}
and constant normal stress is enforced by $\tau_3 = -\sigma_n$, or
\begin{equation}
    -\sigma_n = p_0 + \dfrac{\partial \psi}{\partial {\dot{\theta}^p}}(\dot{\theta}^p, \dot{\eps}^p) + \dfrac{\dot{\theta}^p}{\tilde{\alpha}^2 \dot{\eps}^p} \dfrac{\partial \psi}{\partial \dot{\eps}^p}(\dot{\theta}^p, \dot{\eps}^p),
\label{eq::sigman_gen}
\end{equation}
where $p_0$ and $\psi$ are given by~\eqref{eq::p0_exp}. Solving~\eqref{eq::sigman_gen} effectively provides an evolution equation for ${\theta}^p$, which serves as a state variable analogous to that of rate-and-state friction formulations (e.~g.~\cite{Dieterich1979, Ruina1983, Segall1995}), and corresponds to the volume change necessary to maintain the prescribed normal stress $\sigma_n$ at the plastic strain rate $\dot{\eps}^p$. That is, the derivatives of the stress potential, $\Phi(\dot{\gam})$, give both the stress-strain relation and the evolution law for the state variable, $\theta^p$. 

Specializing for the rate potential in~\eqref{eq::p0_exp},~\eqref{eq::sigman_gen} becomes
\begin{equation} \label{eq::ode_thp}
    \sigma_0 \left( 1 + \tilde{\varphi}(\dot{\eps}^p)\right) - \tilde{\alpha} \sigma_n = \left(\sigma_0 + \sigma_c\varphi'(\dot{\eps}^p)\right) \dfrac{\dot{\theta}^p}{\tilde{\alpha} \dot{\eps}^p} - \sigma_0 \dot{\eps}^p\tilde{\varphi}'(\dot{\eps}^p)\left(\dfrac{\dot{\theta}^p}{\tilde{\alpha} \dot{\eps}^p}\right)^2 .
\end{equation}

\subsubsection{Behavior over constant loading windows}

We demonstrate that~\eqref{eq::ode_thp} results in a decaying exponential dependence of the shear stress with slip, in line with experimental observations from velocity-step tests and previous theoretical work~\cite{Dieterich1979, Ruina1983}. Although it is not possible to derive a closed-form solution for a general rate potential $\tilde{\varphi}(\dot{\eps}^p)$ and $\varphi(\dot{\eps}^p)$, we make a series of simplifying assumptions that result in an explicit solution of~\eqref{eq::ode_thp}. Then, using this solution in~\eqref{eq::tau2_gen} gives the desired result.

To solve~\eqref{eq::ode_thp} explicitly, we assume that the shearing rate is much larger than the dilation rate ($|\dot{\gamma}_2| \gg |\dot{\gamma}_3|$), the volume changes are small ($h \approx h_0$), and the viscous stresses are much smaller than the normal stresses ($\sigma_c\varphi'(\dot{\eps}^p) \ll \tilde{\alpha}\sigma_n$)~(\ref{sec::app_approx_ode}). In this regime,~\eqref{eq::ode_thp} is well-approximated by
\begin{equation} \label{eq::sigma0_evol}
    \dot{\hat{\sigma}}_0 = -\dfrac{\so - \soOc}{t_c},
\end{equation}
where we have used~\eqref{eq::oed_restrict} and~\eqref{eq::p0_exp} to write the ODE only in terms of $\sigma_0$, and
\begin{equation}
    \so := \dfrac{\sigma_0}{\sigma_c}, 
    \quad 
    \dot{\eps}^p \approx \dfrac{1}{\sqrt{3}}\dfrac{\dot{\delta}_2}{h_0}, \quad
    t_c := \dfrac{\theta^p_{\rm ref}/(\tilde{\alpha} \dot{\eps}^p)}{1 + \tilde{\varphi}(\dot{\eps}^p)}, 
    \quad 
    \hat{\sigma}_n := \dfrac{\tilde{\alpha}\sigma_n}{\sigma_c}, 
    \quad 
    \soOc := \dfrac{\hat{\sigma}_n}{1 + \tilde{\varphi}(\dot{\eps}^p)}.
\label{eq::ndim_define}
\end{equation}

For constant loading rate $\dot{\delta}_2$ in the interval $(t_0, t)$ and initial condition $\sigma_0(t_0) = \sigma_0^{\rm init}$, the solution of the ODE~\eqref{eq::sigma0_evol} is given by
\begin{equation}
    \so = \soOc + \left(\soI-\soOc\right)\exp{\left(-\dfrac{t-t_0}{t_c}\right)},
\end{equation}
where, over long loading windows $(t-t_0)/t_c\gg1$, $\so$ evolves to a rate- and normal-stress-dependent steady-state value, $\soOc$. For constant shear rate, time and slip are related through $\dot{\delta_2}$, and thus,~\eqref{eq::sigma0_sol} can be understood as a decaying exponential dependence with slip, or
\begin{equation}\label{eq::sigma0_sol}
    \so = \soOc + \left(\soI-\soOc\right)\exp{\left(-\dfrac{\delta_2-\delta_{20}}{\delta_{2c}}\right)},
\end{equation}
where $\delta_{2c}$ is a rate-dependent characteristic slip distance which follows from substitution into~\eqref{eq::ndim_define},
\begin{equation}
    \delta_{2c} := \dot{\delta}_2 t_c = \dfrac{\sqrt{3}\theta^p_{\rm ref}/\tilde{\alpha}}{1 + \tilde{\varphi}(\dot{\eps}^p)}h_0.
\end{equation}

Using~\eqref{eq::sigma0_sol}, the layer height and shear stress evolutions follow from straightforward substitutions. The change in layer width,~\eqref{eq::h_thp}, is thus
\begin{equation} \label{eq::h_const_ld}
    h = h_0 (2 \so)^{-\theta^p_{\rm ref}},
\end{equation}
where the layer dilates with slip if $\soI >\soOc$, while for $\soI < \soOc$, it compacts. Similarly, the shear stress~\eqref{eq::tau2_gen} is
\begin{equation}\label{eq::tau2}
    \tau_2 = \dfrac{\sigma_c(T)}{\sqrt{3}} \left(\so + \varphi'(\dot{\eps}^p) + \dfrac{\theta^p_{\rm ref}}{\tilde{\alpha}} \tilde{\varphi}'(\dot{\eps}^p) \, \dot{\hat{\sigma}}_0\right),
\end{equation}
where we have used~\eqref{eq::oed_restrict} and~\eqref{eq::p0_exp}. The first term characterizes the change in shear stress with slip: if $\soI >\soOc$, the shear stress decreases with slip and the layer softens, while for $\soI < \soOc$, it increases and the layer hardens. These trends in consolidation-dependent changes in the layer response agree with the experimental results from Figure~\ref{fig::das_1983}. The next two terms describe the rate-dependence of the shear stress, with the last term also describing how far the system is from steady state through $\dot{\hat{\sigma}}_0$.

\subsubsection{Long-term behavior at the critical state}
Laboratory experiments find that a granular sample sheared at a constant loading rate $\dot{\delta}_2$ evolves to a rate-dependent steady-state layer width $h_{\rm crit}$, which increases with the loading rate $\dot{\delta}_2$. Evaluating~\eqref{eq::h_const_ld} in the limit $\so \to \soOc$ gives
\begin{equation} \label{eq::h_crit}
    h_{\rm crit} = h_0\left(\dfrac{1 + \tilde{\varphi}(\dot{\eps}^p)}{2 \hat{\sigma}_n}\right)^{\theta^p_{\rm ref}}.
\end{equation}
From~\eqref{eq::ndim_define}, at constant normal confinement $\sigma_n$, $1/\hat{\sigma}_n$ is smaller for a loosely consolidated material than for a densely consolidated one. When $1/\hat{\sigma}_n>2$, corresponding to more consolidated materials,~\eqref{eq::h_crit} predicts dilation. In addition, for $\tilde{\varphi}$ monotonically increasing, the layer height increases at faster loading rates. Both predictions agree with the key mechanisms from Section~\ref{sec::key_mechanisms}.

At the critical state, the shear stress becomes
\begin{equation} \label{eq::tau_ss}
    \tau_2^{\rm crit} = \dfrac{\tilde{\alpha}\sigma_n}{\sqrt{3}} \left(\left(1 + \tilde{\varphi}(\dot{\eps}^p)\right)^{-1} + \dfrac{1}{\hat{\sigma}_n} \varphi'(\dot{\eps}^p)\right).
\end{equation}
Interestingly, the last term differs from the typical assumption of Coulomb friction, in which the friction coefficient is independent of normal stress. This additional normal stress dependence in~\eqref{eq::tau_ss} occurs when the current confining normal stress is smaller than the pre-consolidation pressure, $\hat{\sigma}_n < 1$. For small $\dot{\eps}^p$, the critical state shear stress is approximately given by
\begin{equation}
    \tau_2^{\rm crit} \approx \dfrac{\tilde{\alpha}\sigma_n}{\sqrt{3}} \left(1 - \tilde{\varphi}(\dot{\eps}^p) + \dfrac{1}{\hat{\sigma}_n}\varphi'(\dot{\eps}^p)\right),
\end{equation}
where the rate effects perturb the friction coefficient around a rate independent value, $\tilde{\alpha}/\sqrt{3}$. Depending on the relative magnitudes of $\tilde{\varphi}$, $\varphi$, and $\hat{\sigma}_n$, $\tau_2^{\rm crit}$ may increase or decrease with loading rate. That is, the response may be \textit{steady-state rate strengthening} or \textit{steady-state rate weakening}. Steady-state rate strengthening results in stable, slow slip under tectonic plate loading, while the weakening of the interface with larger loading rates leads to dynamic ruptures, or earthquakes \cite{Rice1983, Lapusta2000}; hence, the ability of the model to capture both behaviors could explain the existence of both stably slipping and stick-slip fault segments.  

\subsection{Specialization of rate potentials}
To illustrate the behavior of the model, we select functional forms for the rate potentials $\tilde{\varphi}$ and $\varphi$ satisfying the convexity constraints from~\ref{sec::app_convexity}. We consider two examples,
\begin{equation}
    \begin{aligned}
        &\text{Case 1: } \quad \varphi(\dot{\eps}^p)  = \eta_1\dfrac{\dot{\eps}^p_0}{2}\left(\dfrac{\dot{\eps}^p}{\dot{\eps}^p_0}\right)^2, \quad &\tilde{\varphi}(\dot{\eps}^p) = \frac{\eta_2}{\eta_1}\dfrac{2 \varphi(\dot{\eps}^p)}{ \dot{\eps}^p},\\
        &\text{Case 2: }\quad  \varphi(\dot{\eps}^p)  = \eta_1 \dot{\eps}^p_0\left(1-\sqrt{1 + \left(\dfrac{\dot{\eps}^p}{\dot{\eps}^p_0}\right)^2} + \dfrac{\dot{\eps}^p}{\dot{\eps}^p_0} \sinh^{-1}\left(\dfrac{\dot{\eps}^p}{\dot{\eps}^p_0}\right)\right), \quad &\tilde{\varphi}(\dot{\eps}^p) = \frac{\eta_2}{\eta_1} \dfrac{2 \varphi(\dot{\eps}^p)}{\dot{\eps}^p},
    \end{aligned}
    \label{eq::rateSensitivities}
\end{equation}
where the first case corresponds to linear rate-sensitivity and the second is motivated by activation theory~\cite{Rice2001}. For small plastic strain rates, $\dot{\eps}^p \ll \dot{\eps}^p_0$, the two rate potentials are identical at leading order. In the following section, we conduct numerical experiments and compare the response of the material model for each case of rate sensitivity.

\section{Model validation} \label{sec::results}

We investigate the response of the material model to tests commonly used to characterize granular media. First, we conduct a numerical oedometer test, in which we increase the normal load at a constant loading rate, $\dot{\sigma}_{ld}$, until reaching the maximum load, $\sigma_M$. Next, we simulate a shear test with constant and stepped loading and examine the volumetric and shear stress responses. In both tests, we explore the changes in material response associated with changes in the confinement to consolidation ratio $\hat{\sigma}_n$. 

To validate the model, we solve the stress-strain relations~\eqref{eq::sig_shearlayer} numerically at a material point. We first discretize the equations using an implicit backward-Euler scheme. We then write the layer height, plastic strain rate, and state variable in terms of the slip increments, $\Delta \delta_i$, for $i = 1,2,3$. Next, the stresses at each timestep for displacement-controlled deformations follow from substitution. For load-controlled deformations, we solve for the slip increments using Newton-Raphson iteration. The details of the numerical solution procedure are described in~\ref{sec::app_numerics}.

\subsection{Oedometer test}
\label{sec::results_oed}
In the simulated oedometer test, we prescribe the lateral confinement, $\delta_1 = \delta_2 = 0$, and the vertical compressive stress loading rate, $\dot{\sigma}_{ld}$, and compress the material to a maximum vertical stress $\sigma_M$ before unloading. We probe the effect of consolidation on the stress-strain response by modifying the normalized pre-consolidation stress, $p_c/\sigma_M$. 
From~\eqref{eq::consol_test}, we expect increasing pre-consolidation stress to result in a stiffer response.

\begin{figure}[t!]
    \centering
    \includegraphics[width=0.95\linewidth]{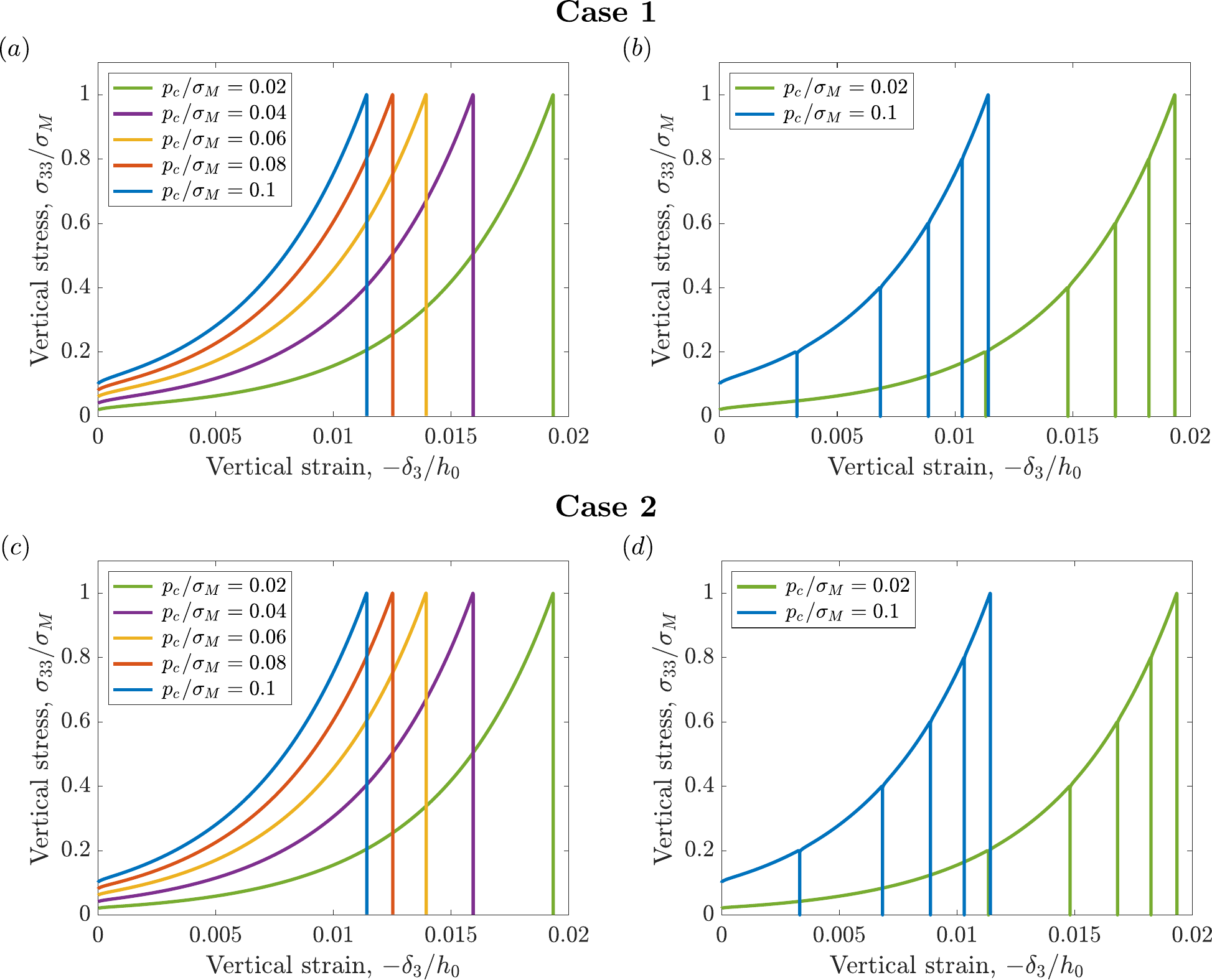}
    \caption{Numerical cyclic oedometer tests for $\dot{\eps}^p_0 = 10^{-5} \, \rm s^{-1}$ and two different types of rate sensitivity. (a) and (b) Linear rate sensitivity (Case 1), with (a) one and (b) five loading-unloading cycles. (c) and (d) Activation-motivated rate sensitivity (Case 2), with (c) one and (d) five loading-unloading cycles. In this parameter regime, both types of rate sensitivity predict similar responses. Increasing consolidation stiffens the material response, while the assumption of rigid elasticity leads to vertical unloading curves. Cyclic loading at different consolidations highlights the increasing stiffness at larger strains. At large strains, the effect of pre-consolidation on the tangent stiffness is negligible.
}
    \label{fig::num_oed_test_1}
\end{figure}

Figure~\ref{fig::num_oed_test_1} shows the oedometer tests for a reference strain rate $\dot{\eps}^p_0 = 10^{-5} \, \rm s^{-1}$, with the remaining plasticity parameters shown in Table~\ref{tab::oed_params}. The top and bottom rows compare the two rate sensitivity cases defined in \eqref{eq::rateSensitivities}, while the left and right columns correspond to single and multiple loading-unloading cycles. In this parameter regime, the response is identical for both linear and activation-motivated rate sensitivities. The oedometric behavior in response to a single loading-unloading cycle, shown in Figures~\ref{fig::num_oed_test_1}(a) and~\ref{fig::num_oed_test_1}(c), is characterized by stiffening with increasing pre-consolidation, increasing tangent modulus during loading, and rigid-elastic unloading.

\begin{table}[t]
\begin{center}
\begin{tabular}{|p{2.25cm}|| p{2.75cm}|p{7cm}|} 
 \hline
 \multicolumn{3}{|c|}{\textbf{Simulation parameters}} \\
 \hline
 Symbol & Value & Description \\ 
 \hline
 $\alpha$ & $1.2$ & Triaxial internal friction coefficient \\
 $\theta^p_{\rm ref}$ & $0.005$ & Reference volumetric strain \\
 $\eta_1$ & $0.015$ & Rate-potential $\varphi$ coefficient \\
 $\eta_2$ & $0.025$ & Rate-potential $\tilde{\varphi}$ coefficient \\
 $\dot{\eps}^p_0$ & $10^{-5}$ s$^{-1}$ & Reference plastic strain rate \\
 \hline
 \multicolumn{3}{|c|}{\textbf{Oedometer test}} \\
 \hline
 $\dot{\sigma}_{ld}/\sigma_M$ & $2 \, \rm{h}^{-1}$ & Compressive loading rate \\
 $\dot{\eps}^{p,\, \rm mod}_0$ & $10^{-7}$ s$^{-1}$ & Modified reference plastic strain rate \\
 \hline
 \multicolumn{3}{|c|}{\textbf{Shear test}} \\
 \hline
 $\dot{\delta}_2^i/h_0$ & $10^{-5}$ s$^{-1}$ & Initial (slower) shear loading rate \\
 $\dot{\delta}_2^j/h_0$ & $10^{-4}$ s$^{-1}$ & Faster shear loading rate \\
 \hline
\end{tabular}
\end{center}
\caption{Simulation parameters for examples.}
\label{tab::oed_params}
\end{table}

We compare the effect of pre-consolidation on the model response to cyclic loading in Figures~\ref{fig::num_oed_test_1}(b) and~\ref{fig::num_oed_test_1}(d). In the first loading-unloading cycle, the less consolidated material has a softer response, reaching larger vertical strains for the same stress in comparison to the more consolidated material. However, the effect of pre-consolidation diminishes in the subsequent cycles as the material gets loaded to vertical stresses much larger than the pre-consolidation pressure. This corresponds to the first term of~\eqref{eq::consol_test} becoming much larger than the second term as the material becomes more consolidated.

\begin{figure}[t!]
    \centering
    \includegraphics[width=0.95\linewidth]{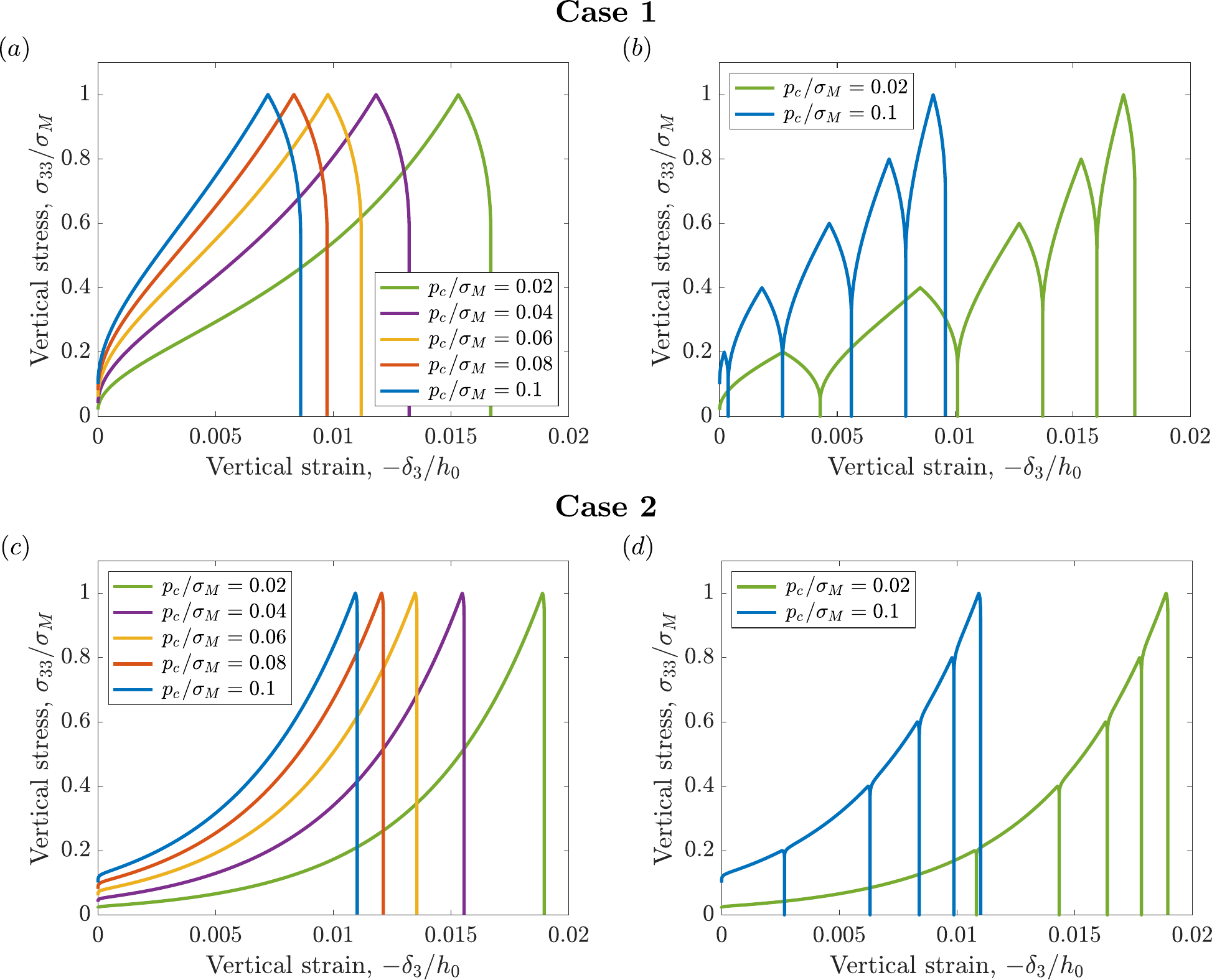}
    \caption{Numerical cyclic oedometer tests for $\dot{\eps}^{p, \, \rm mod}_0 = 10^{-7} \, \rm s^{-1}$ and two different types of rate sensitivity. 
    (a) and (b) Linear rate sensitivity (Case 1), with (a) one and (b) five loading-unloading cycles. (c) and (d) Activation-motivated rate sensitivity (Case 2), with (c) one and (d) five loading-unloading cycles. Decreasing $\dot{\eps}^p_0$ highlights the differences in the rate sensitivity at strain rates much larger than the reference value. }
    \label{fig::num_oed_test_2}
\end{figure}

The two cases of rate sensitivity are nearly identical at leading order for $\dot{\eps}^p$ comparable to $\dot{\eps}^p_0$. To highlight their differences in the regime $\dot{\eps}^p \gg \dot{\eps}^p_0$, we decrease the reference plastic strain rate by two orders of magnitude to $\dot{\eps}^{p,\, \rm mod}_0 = 10^{-7}~{\rm s}^{-1}$. The resulting oedometer tests are shown in Figure~\ref{fig::num_oed_test_2}. Although the effect of pre-consolidation is qualitatively similar to the previous example, the stress-strain curves for the two rate sensitivity cases have marked differences. 

In the new parameter regime, the response is much stiffer at small vertical strain for the linear rate sensitivity (Figure~\ref{fig::num_oed_test_2}(a) and~\ref{fig::num_oed_test_2}(b)). Additionally, there is a transient viscous response during unloading. In comparison, the response for the activation-motivated rate potential, Figure~\ref{fig::num_oed_test_2}(c) and~\ref{fig::num_oed_test_2}(d), looks similar in both parameter regimes. The difference in response arises due to differences in $\varphi'$ between the rate sensitivity examples at large plastic strain rates. In response to cyclic loading, the linear case, Figure~\ref{fig::num_oed_test_2}(b), is much stiffer in the first cycle, but recovers final strains similar to the activation-motivated case, Figure~\ref{fig::num_oed_test_2}(d), after multiple loading cycles.

The numerical results reported in Figures~\ref{fig::num_oed_test_1} and~\ref{fig::num_oed_test_2} qualitatively agree with the experimental results shown in Figure~\ref{fig::brzesowsky_oed_test}. Our model predicts the same trend of stiffening with increasing consolidation observed experimentally for both cases of rate sensitivity considered. Slight differences arise in the unloading response as a result of our assumption of rigid elasticity. Otherwise, we find excellent agreement between our model and the laboratory oedometer tests.

\subsection{Shear test}
Next, we simulate a plane-strain shear test with $\delta_1 = 0$ and $\delta_2(t)>0$ prescribed through constant and stepped loading rates, using the same plasticity parameters as for the oedometer test in Section~\ref{sec::results_oed}. We investigate the role of pre-consolidation and confining pressure on the material's response in shear by modifying the confinement to pre-consolidation ratio, $\hat{\sigma}_n$. For the constant loading rate test, we view the changes in $\hat{\sigma}_n$ as changes in pre-consolidation for a fixed normal stress to directly compare to the experimental results from Figure~\ref{fig::das_1983}. In contrast, for the stepped loading rate example, we view the changes in $\hat{\sigma}_n$ as changes in normal stress for a fixed pre-consolidation stress to draw comparisons to laboratory friction experiments.

\subsubsection{Constant loading rate}
The schematic in Figure~\ref{fig::das_1983} demonstrates the effects of pre-consolidation on shear layers subject to constant shearing rate and normal stress; granular materials dilate and soften when highly consolidated and compact and harden when loosely consolidated. In our material model, this difference in response corresponds to stress states on opposite sides of the critical state line in Figure~\ref{fig::yield_surf}, with the transition point around $1/\hat{\sigma}_n =~2$~(\ref{sec::app_sigman_hat}).

\begin{figure}[b!]
    \centering
    \includegraphics[width=0.8\linewidth]{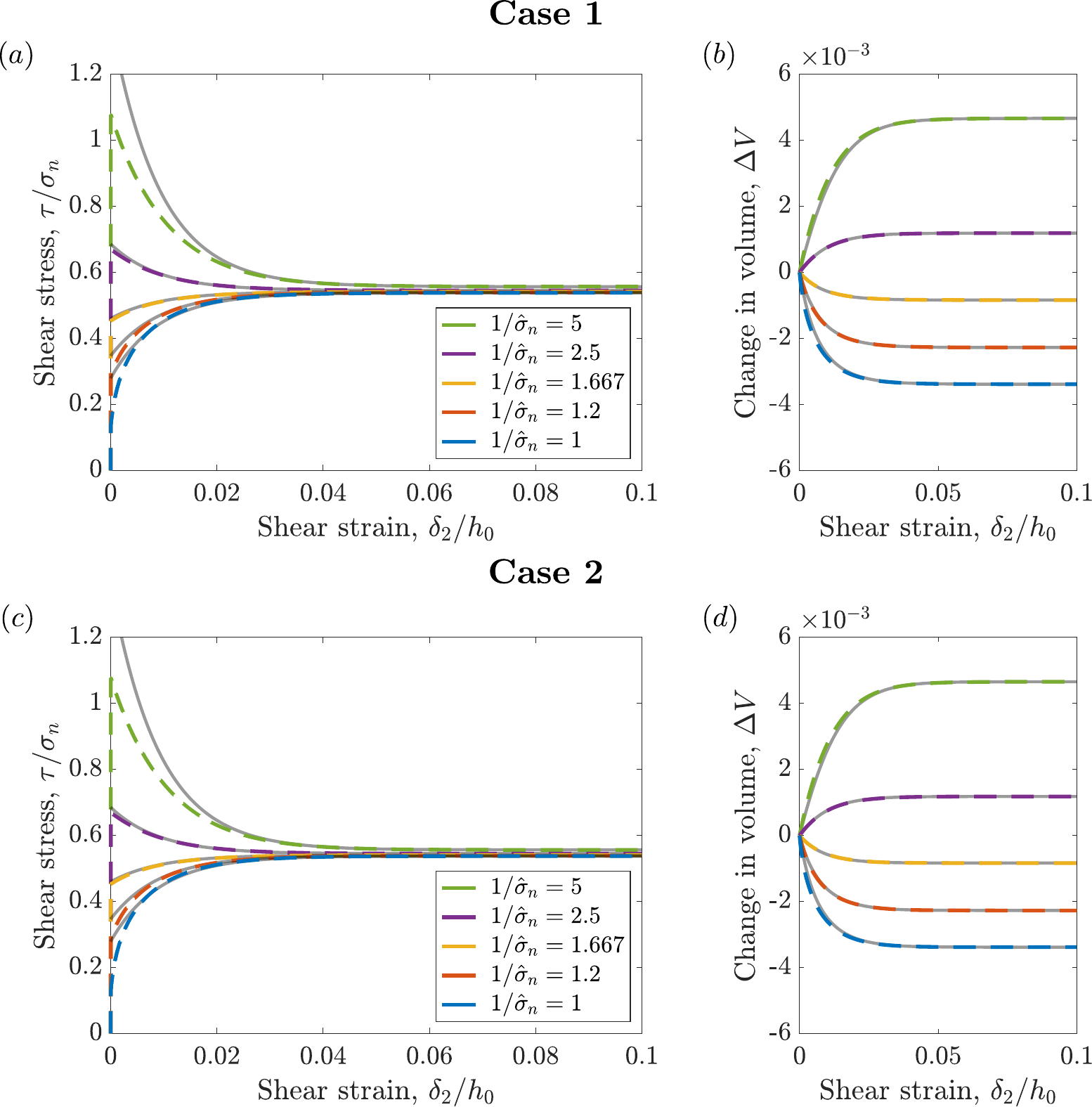}
    \caption{Numerical response of a granular layer to constant shear loading, with varying consolidation pressure and constant normal stress, and its comparison to theoretical predictions, for two rate sensitivity cases. Dashed lines correspond to numerical solutions and solid lines are predictions from~\eqref{eq::tau2} and~\eqref{eq::h_const_ld} with~\eqref{eq::sigma0_sol}. (a) and (b) Linear rate sensitivity (Case 1), with (a) shear stress and (b) volumetric strain versus shear strain. (c) and (d) Activation-motivated rate sensitivity (Case 2), with (c) shear stress and (d) volumetric strain versus shear strain. The rate sensitivity cases behave similarly. There is a transition from hardening and compaction to softening and dilation with increasing consolidation $1/\hat{\sigma}_n$.} 
    \label{fig::const_shear}
\end{figure}

To study the role of pre-consolidation in the shear stress and volumetric response of the material model, we increase $\hat{\sigma}_n$ and compare to the theoretical predictions from Section~\ref{sec::shear_test} (Figure~\ref{fig::const_shear}). The initial elastic response seen in Figure~\ref{fig::das_1983} is not recovered due to our assumption of rigid elasticity. However, for both rate-sensitivity examples, we find that the results qualitatively agree at large strains; above $1/\hat{\sigma}_n = 2$, there is a dilatant, softening response characteristic of highly consolidated granular materials, and for $1/\hat{\sigma}_n < 2$, the response changes to that of loosely consolidated samples. At large shear strains, the steady-state friction coefficient $\tau/\sigma_n \approx 0.6$ for all $\hat{\sigma}_n$, consistent with Byerlee friction~\cite{Byerlee1978} (Figures~\ref{fig::const_shear}(a) and~\ref{fig::const_shear}(c)). In contrast, the steady-state layer heights, Figures~\ref{fig::const_shear}(b) and~\ref{fig::const_shear}(d), are consolidation-dependent; larger deviations from $1/\hat{\sigma}_n = 2$ result in increased absolute changes in layer height. For both types of rate sensitivity, there is good agreement between the numerical results and the theoretical predictions for the shear stresses~\eqref{eq::tau2} and layer height~\eqref{eq::h_const_ld}.

\subsubsection{Stepped loading rate}
\label{sec::step_test}
\begin{figure}[htb!]
    \centering
    \includegraphics[width=0.8\linewidth]{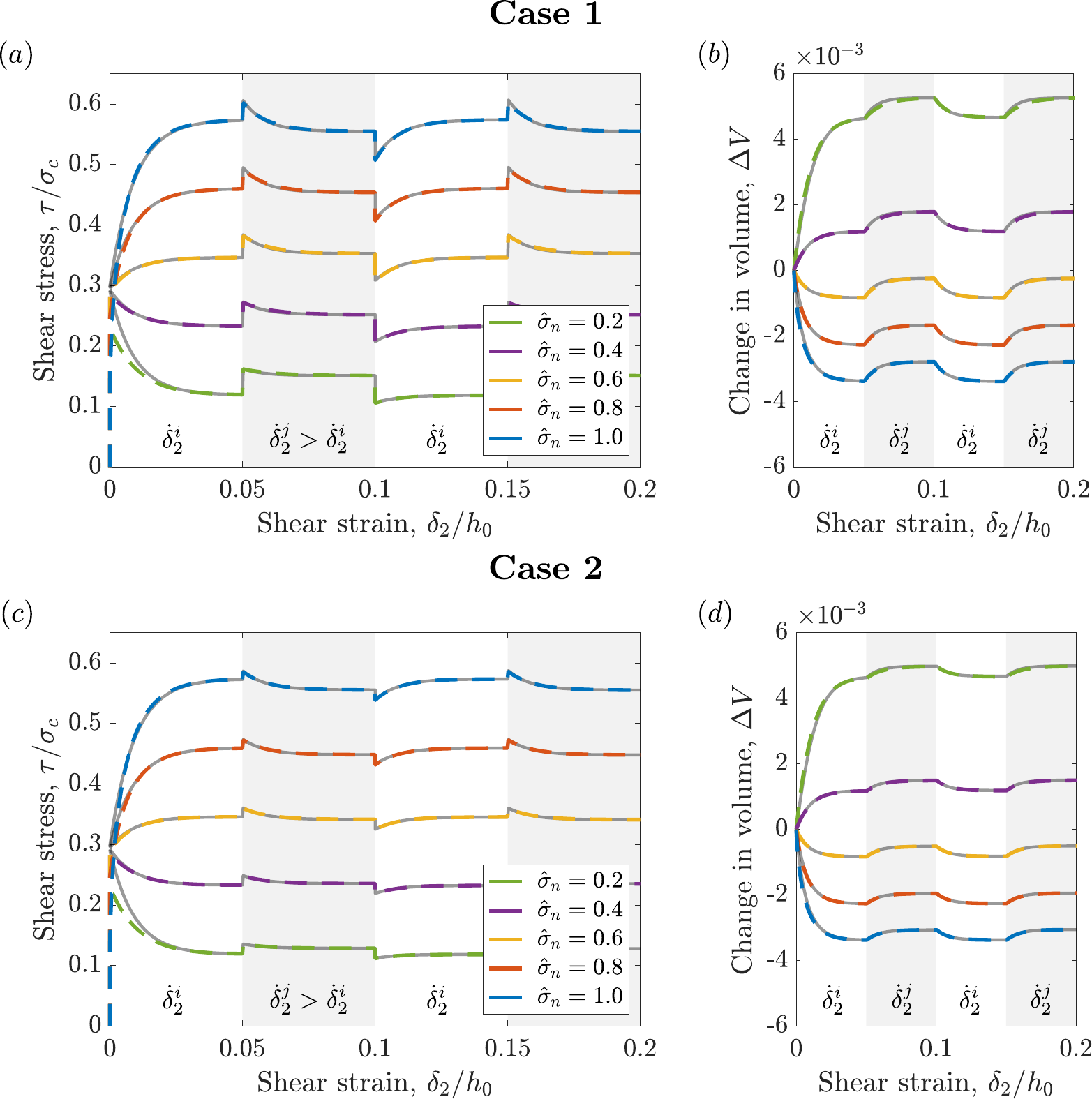}
    \caption{Numerical response of a granular layer to stepped shear loading, with varying normal stress and constant consolidation pressure, and its comparison to theoretical predictions, for two rate sensitivity cases. The shear displacement rate is changed every $5\%$ strain. Dashed lines correspond to numerical solutions and solid lines are theoretical predictions from~\eqref{eq::tau2} and~\eqref{eq::h_const_ld} with~\eqref{eq::sigma0_sol}. (a) and (b) Linear rate sensitivity (Case 1), with (a) shear stress and (b) volumetric strain versus shear strain. (c) and (d) Activation-motivated rate sensitivity (Case 2), with (c) shear stress and (d) volumetric strain versus shear strain. In response to a step-change in slip rate, there is a rate-dependent change in shear stress followed by a slip-dependent transition to a new steady state. The volume increases at faster loading rates. The linear rate sensitivity exhibits stronger rate-dependence than the activation-motivated rate sensitivity. }
    \label{fig::step_shear}
\end{figure}

The most common material characterization test in experimental geomechanics is the shear velocity-step test~\cite{Dieterich1979, Ruina1983, marone1990frictional, Bedford2022}. In this test, the normal stress is typically held constant, and the shear displacement rate is alternatively held constant and changed in abrupt step increments. The shear stress and volumetric response is recorded. The typical experimental response is shown in Figure~\ref{fig::rathbun_marone}. We compare the experimental results to the numerical tests for different $\hat{\sigma}_n$ in Figure~\ref{fig::step_shear}. Here, we attribute changes in $\hat{\sigma}_n$ to changes in normal stress for a fixed pre-consolidation stress. We note that the shear stress in Figure~\ref{fig::step_shear} is nondimensionalized by the pre-consolidation stress and not the normal stress. As a result, the vertical axis in Figures~\ref{fig::step_shear}(a) and~\ref{fig::step_shear}(c) is not the friction coefficient; the friction coefficient at steady state is close to 0.6 for all cases shown.

The shear stresses qualitatively agree with the experimental results. There is an immediate change in shear stress in response to a step change in displacement rate, with a larger stress change for the linear rate sensitivity case, Figure~\ref{fig::step_shear}(a), than for the activation-motivated rate sensitivity case, Figure~\ref{fig::step_shear}(c). From~\eqref{eq::tau2}, the difference in response occurs as a result of differences in $\varphi'$ between the rate sensitivities -- i.~e. the slower growth of $\sinh^{-1}$ compared to the linear case at large arguments. 

\begin{figure}
    \centering
    \includegraphics[width=0.95\linewidth]{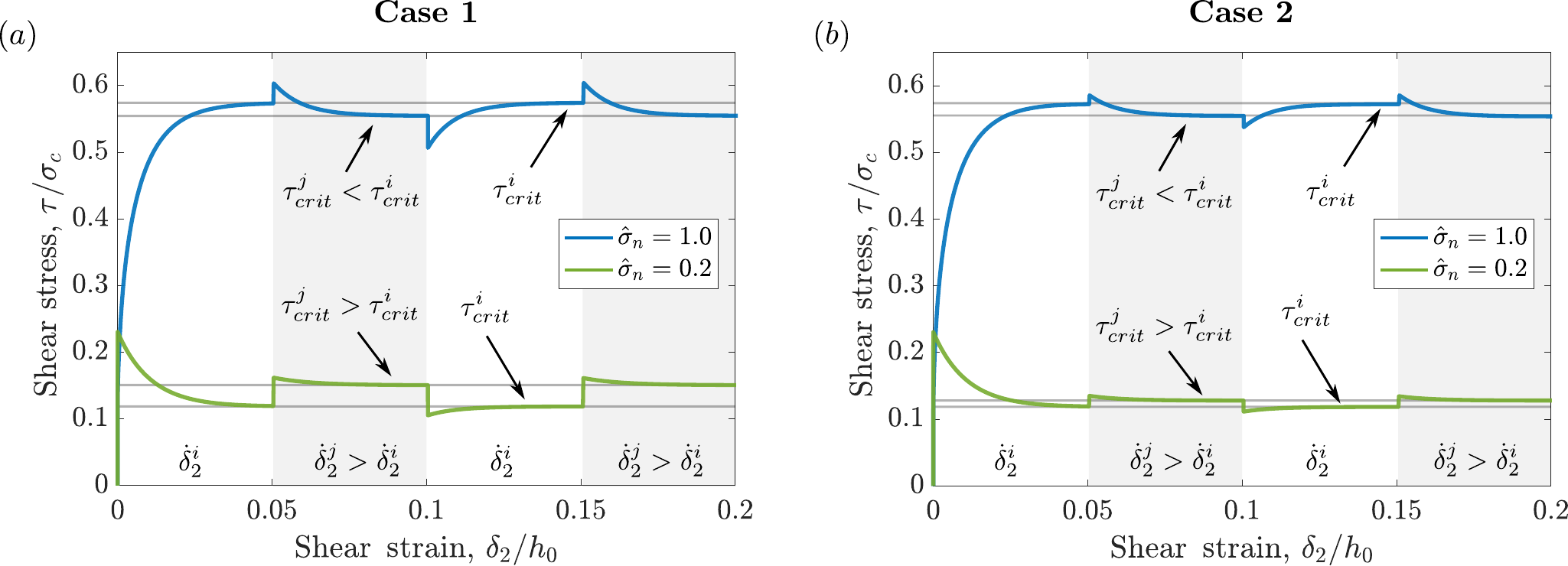}
    \caption{Comparison of steady-state rate-dependent shear stress from Figure~\ref{fig::step_shear} for two rate sensitivity cases. For $\hat{\sigma}_n = 0.2$, the critical-state shear stresses are larger at the faster loading rate. The opposite is true for $\hat{\sigma}_n = 1.0$, where the critical-state shear stresses are smaller at the faster loading rate. This indicates a transition from steady-state rate-strengthening to steady-state rate-weakening with increasing $\hat{\sigma}_n$.}
    \label{fig::ss_rate_comp}
\end{figure}

After the initial rate-dependent response, both rate sensitivity cases transition to a rate-dependent steady-state stress with shear strain, as observed in the constant loading rate tests and laboratory experiments. Notably, at the steady-state, there is a transition from steady-state rate strengthening to steady-state rate weakening with increasing $\hat{\sigma}_n$ consistent with~\eqref{eq::tau_ss} (Figure~\ref{fig::ss_rate_comp}). While the change in rheology is more dramatic for the linear rate sensitivity case, Figure~\ref{fig::ss_rate_comp}(a), than the activation-motivated case, Figure~\ref{fig::ss_rate_comp}(b), both examples demonstrate that similarly consolidated granular materials become more unstable at higher confinements $\hat{\sigma}_n$. This change in the steady-state layer behavior with $\hat{\sigma}_n$ could lead to faults experiencing stable creep and seismic slip in the same locations. 

We next compare the change in volume for the two rate-sensitivity cases, Figures~\ref{fig::step_shear}(b) and~\ref{fig::step_shear}(d), to the experimental findings, Figure~\ref{fig::rathbun_marone}(b). The material model exhibits dilation at faster loading rates for both rate sensitivity cases, with larger changes in volume for linear rate sensitivity, Figure~\ref{fig::step_shear}(b), than the activation-motivated rate sensitivity, Figure~\ref{fig::step_shear}(d). These trends agree with the theoretic prediction~\eqref{eq::h_crit} and the experimental results. 

In Section~\ref{sec::shear_test}, we made simplifying assumptions in order to derive explicit solutions for the shear stress and state variable evolution. Despite these approximations, we observe excellent agreement between the linearized theoretical predictions and the numerical results for both the transient and steady-state shear stress and layer height. This validates our assumptions from~\ref{sec::app_approx_ode} for the parameter regime considered.

We conduct additional parameter studies to probe the response of the model in~\ref{sec::app_step_exs}. The examples illustrate the behavior of the model in response to a larger step in shear rate, $\dot{\delta}_2^j/\dot{\delta}_2^i = 100$, and stronger rate weakening, $\eta_2 = 0.035$. The results are congruous with the analysis presented in the main text.

\newpage
\section{Conclusions}
\label{sec:conclusions}

We have developed a variational, finite-deformation framework for non-isochoric, rigid-plastic materials such as fault gouge. The variational structure of the governing equations admits a potential formulation for the effective stress–strain relations. In the particular case of shearing and dilating granular layers, the shear stresses and evolution of internal variables are prescribed by the choice of the rate potential. By invoking a separation of scales between the shear and dilatant responses, we obtain closed-form expressions for the shear stress and layer dilation in simple shear. Lastly, we demonstrate the ability of the model and theory to capture experimental features from consolidation and shear tests.

Our formulation extends existing descriptions of granular material behavior in several important respects. First, the rate and direction of plastic flow follow from the maximization of a dissipation potential subject to a kinematic constraint. The governing equations are cast in rate form and are therefore independent of numerical discretization. The variational structure guarantees objectivity and provides a consistent extension of yield surfaces to finite deformations. In addition, the effective stress–strain relations emerge from the stationarity of the dissipation functional. This potential-based formulation thus resolves ambiguities concerning the evolution of internal variables for shearing interfaces and offers a clear physical interpretation of the internal variable as the effective consolidation state of the layer. A further advantage of the approach lies in its calibration procedure, which requires only standard consolidation experiments and convexity considerations. Consequently, the predicted response to standard shearing protocols follows naturally from the variational structure.

The model reproduces key qualitative features of experimental observations on shearing granular layers. Using two canonical examples of rate sensitivity in viscoplastic theories, we capture the non-monotonic response of simulated fault gouge under stepped shear loading. The analysis identifies the ratio of pre-consolidation to confinement stress, denoted by $\hat{\sigma}_n$, as a principal parameter controlling the shear response. Increasing $\hat{\sigma}_n$ induces a transition in the rheological behavior, leading to rate weakening—a destabilizing mechanism naturally associated with earthquake nucleation in fault zones.

The proposed framework introduces significant improvements to the empirical rate-and-state description of friction, a model known for reproducing a wide range of earthquake source observations with a minimal parameter set~\cite{Lapusta2000, Rubino2017, scholz2019mechanics}. The rate-and-state parameters are typically inferred only from shear tests and are known to vary with loading history, normal stress, and other factors~\cite{marone1990frictional, Beeler1994, Beeler1996, mair1999friction}, undermining their interpretation as material constants. Additionally, although the effect of dilatancy has been incorporated into rate-and-state friction formulations (e.~g. \citet{Segall1995, Sleep1995}), it has been done on the basis of a limited set of experiments. 

The present variational formulation offers a rational foundation for these empirical laws, provides a framework for including a wealth of additional information to determine the appropriate functional forms and parameters for the shear resistance of granular layers, and suggests a path toward identifying consistent analogues between rate-and-state parameters and the critical-state variables of our model. Systematic comparisons of the developed framework with laboratory experiments under diverse loading protocols, for different minerological compositions of gouge, and at the faster strain-rates characteristic of earthquakes that activate shear-induced heating and hence temperature effects (e.~g.~\cite{Tsutsumi1997, Rice2006, DiToro2011, Tullis2015,Rubino2017}) constitute a natural continuation of this work.

Finally, the extension of the framework to account for pore fluid effects offers a promising direction. In fluid-permeated fault zones, pore pressure reduces the effective normal stress and thereby facilitates slip. Conventional friction models incorporate this effect via the Terzaghi effective stress, wherein the normal stress is reduced by the pore pressure. This modification arises naturally in our setting through a straightforward adjustment of the dissipation function. More elaborate couplings linking volumetric deformation, pore pressure evolution, and fluid transport would further enhance the model’s fidelity and provide valuable insight into the mechanics of fluid-assisted fault slip.

\section*{Acknowledgments}

N.L. and M.A. gratefully acknowledge support from the National Science Foundation (grant EAR 2139331). M.O. gratefully acknowledges the financial support of the Centre Internacional de M\`etodes Num\`erics a l’Enginyeria (CIMNE) of the Universitat Polit\`ecnica de Catalunya (UPC), Spain, through the UNESCO Chair in Numerical Methods in Engineering. A.P. is grateful for the support of the Italian National Group of Physics-Mathematics (GNFM) of the Italian National Institution of High Mathematics ``Francesco Severi'' (INDAM). We thank Kaushik Bhattacharya and Andrew Akerson for insightful discussions and feedback on the manuscript and Ahmed Elbanna for initial conversations about plasticity models for fault zones.

\section*{Data availability}
Data will be made available on request.

\section*{Declaration of competing interest}
The authors declare there are no conflicts of interest for this manuscript.

\newpage
\appendix
\section{Convexity of rate potential}
\label{sec::app_convexity}
We identify constraints on the rate functions $\tilde{\varphi}$ and $\varphi$ for which the rate potential given in Section~\ref{sec::restricted} is convex. A straightforward calculation of the second derivative of the rate potential gives
\begin{equation}
    \dfrac{d^2 \psi}{d \dot{\eps}^{p \, 2}}(\dot{\eps}^p \widebar{M}, \dot{\eps}^p) 
    = 
    \widebar{M}^2 \dfrac{\partial^2 \psi}{\partial \dot{\theta}^{p \, 2}} 
    + 
    \widebar{M} \dfrac{\partial^2 \psi}{\partial \dot{\theta}^p \partial \dot{\eps}^p} 
    + 
    \dfrac{\partial^2 \psi}{\partial \dot{\eps}^{p \, 2}},
\end{equation}
or substituting~\eqref{eq::p0_exp},
\begin{equation}
    \dfrac{d^2 \psi}{d \dot{\eps}^{p \, 2}}(\dot{\eps}^p \widebar{M}, \dot{\eps}^p) = \sigma_c\varphi''(\dot{\eps}^p) - \sigma_0 \left(\dot{\eps}^p\tilde{\varphi}''(\dot{\eps}^p) + 2 \tilde{\varphi}'(\dot{\eps}^p)\right)\frac{\widebar{M}}{\tilde{\alpha}},
\end{equation}
for a fixed $\bM$. Then, convexity is ensured by the conditions
\begin{equation} \label{eq::convex_gen}
    \sigma_c\geq 0, \quad \text{and} \quad 2\varphi'' - (\dot{\eps}^p \tilde{\varphi}'' + 2\tilde{\varphi}') \dfrac{\dot{\theta}^p}{\tilde{\alpha}\dot{\eps}^p}e^{-\theta^p/\theta^p_{\rm ref}} \geq 0,
\end{equation}
where we have used~\eqref{eq::flow_thp}. The condition~\eqref{eq::convex_gen} depends on both the first and second derivatives of $\tilde{\varphi}$, and it is difficult to identify the conditions under which convexity of $\psi(\cdot, \cdot)$ is guaranteed. To overcome this limitation, we suppose the rate potentials $\tilde{\varphi}$ and $\varphi$ satisfy
\begin{equation}
    \tilde{\varphi}(\dot{\eps}^p) = \frac{2}{\eta} \dfrac{\varphi(\dot{\eps}^p)}{\dot{\eps}^p},
\end{equation}
in which case~\eqref{eq::convex_gen} has the simplified form
\begin{equation} \label{eq::convex_1}
    \sigma_c\geq 0, \quad \varphi'' \geq 0, \quad  \text{and} \quad \eta \geq \dfrac{\dot{\theta}^p}{\tilde{\alpha} \dot{\eps}^p}e^{-\theta^p/\theta^p_{\rm ref}}.
\end{equation}

In compaction, $\dot{\theta}^p < 0$, $\psi$ is convex for all $\eta \geq 0$. For the dilatant case, $\dot{\theta}^p > 0$, we bound the plastic strain rate ratio using~\eqref{eq::ep_shearlayer},
\begin{equation} \label{eq::pl_strain_ratio}
    \dfrac{\dot{\theta}^p}{\tilde{\alpha}\dot{\eps}^p} \leq 1.
\end{equation}
In the absence of shear, $\dot{\gamma}_1 = \dot{\gamma}_2 = 0$, the total plastic strain rate $\dot{\eps}^p$ is entirely volumetric, and the equality is achieved. Thus, for volumetric dilatational deformations with $\dot{\theta}^p(t)>0$ and $\dot{\gamma}_1 = \dot{\gamma}_2 = 0$ for all $t$, we require $\eta \geq 1$ for convexity. This condition is overly restrictive for dilatant shear deformations, where the plastic strain rate ratio~\eqref{eq::pl_strain_ratio} may be small (i.~e. due to scale separation between shear and volumetric strains).

\section{Simplification of internal variable evolution law}
\label{sec::app_approx_ode}
We outline the assumptions that allow~\eqref{eq::ode_thp} to be solved explicitly. We first assume that the shearing rate is much larger than the dilation rate such that 
$|\dot{\gamma}_2| \gg |\dot{\gamma}_3|$, or
\begin{equation} \label{eq::lin_ode_thp}
    \sigma_0 \Big( 1 + \tilde{\varphi}(\dot{\eps}^p) + O((\dot{\gamma}_3/\dot{\gamma}_2)^2)\Big) - \tilde{\alpha} \sigma_n = \left(\sigma_0 + \sigma_c\varphi'(\dot{\eps}^p)\right) \dfrac{\dot{\theta}^p}{\tilde{\alpha} \dot{\eps}^p},
\end{equation}
and, at leading order, the plastic strain rate is
\begin{equation}
    \dot{\eps}^p(\dot{\gamma}_2) 
    = 
    \dfrac{\dot{\gamma}_2}{\sqrt{3}}
    \left(1 + O((\dot{\gamma}_3/\dot{\gamma}_2)^2)\right).
\label{eq::def_epdots}
\end{equation}
We further assume that the change in $\dot{\eps}^p$ due to volumetric deformation is small, such that the plastic strain rate can be written using the Lagrangian description
\begin{equation}
    \dot{\eps}^p \approx \dfrac{1}{\sqrt{3}}\dfrac{\dot{\delta}_2}{h_0}.
\label{eq::approx_epdots}
\end{equation}
Over time intervals $(t_0, t)$ of constant loading rate $\dot{\delta}_2$ and with initial condition $\sigma_0(t_0) = \sigma_0^{\rm init}$, the solution of the ODE~\eqref{eq::lin_ode_thp} is then given implicitly by
\begin{equation}
    -\dfrac{t-t_0}{t_c} = \left(1 + \beta\right)\log \left(\dfrac{\so-\soOc}{\soI-\soOc} \right) - \beta \log \dfrac{\so}{\soI},
\label{eq::ode_fullsol}
\end{equation}
where
\begin{equation}
    \so := \dfrac{\sigma_0}{\sigma_c}, 
    \quad 
    t_c := \dfrac{\theta^p_{\rm ref}/(\tilde{\alpha} \dot{\eps}^p)}{1 + \tilde{\varphi}(\dot{\eps}^p)}, 
    \quad 
    \hat{\sigma}_n := \dfrac{\tilde{\alpha}\sigma_n}{\sigma_c}, 
    \quad 
    \soOc := \dfrac{\hat{\sigma}_n}{1 + \tilde{\varphi}(\dot{\eps}^p)}, 
    \quad 
    \beta := \dfrac{\varphi'(\dot{\eps}^p)}{\soOc},
\end{equation}
are the non-dimensionalized state variable, characteristic time, normal stress to pre-consolidation ratio, steady-state state variable for a given shearing rate, and viscous stress to consolidation ratio, respectively. To write~\eqref{eq::ode_fullsol} explicitly for $\so$, we consider the relative magnitude of $\beta$. For $\varphi$ strongly convex, that is, $\varphi'' \geq \xi > 0$ for all $\dot{\eps}^p > 0$, $\beta$ is small when the viscous effects are small compared to the confinement, or when the strain rates satisfy
\begin{equation}
    \dfrac{\xi \, \dot{\eps}^p}{\soOc} \ll 1, \quad \text{or} \quad
    \dot{\eps}^p \ll \dfrac{\hat{\sigma}_n}{\xi}.
\end{equation}
In this regime, the ODE and its solution are well-approximated by
\begin{equation}
    \dot{\hat{\sigma}}_0 = -\dfrac{\so - \soOc}{t_c}, 
    \quad 
    \so = \soOc + \left(\soI-\soOc\right)\exp{\left(-\dfrac{t-t_0}{t_c}\right)}.
\end{equation}

\section{Numerical solution procedure}
\label{sec::app_numerics}
We describe below the numerical solution procedure used to conduct the model validation tests in Section~\ref{sec::results}. The stress-strain relations,~\eqref{eq::sig_shearlayer}, are written incrementally using a backward-Euler scheme and solved numerically at a material point. We begin by discretizing the layer width,~\eqref{eq::h_thp}, and plastic strain rate,~\eqref{eq::ep_shearlayer}, at the next timestep,
\begin{subequations}
\begin{align}
    &h_{n+1} = h_n + \Delta \delta_3, \\
    &\Delta \eps^{p \, 2} = \frac{1}{h^2_{n+1}}\left(\frac{\Delta \delta_1^2}{3} + \frac{\Delta \delta_2^2}{3} + \frac{\Delta \delta_3^2}{\tilde{\alpha}^2}\right), \\
    &\Delta \theta^p = \frac{\Delta \delta_3}{h_{n+1}},
\end{align}
\label{eq::discrete_eps}
\end{subequations}
such that the governing equations are written in terms of the slip increments. Then, inserting the discretization~\eqref{eq::discrete_eps} into~\eqref{eq::sig_shearlayer} and evaluating the state-dependent functions implicitly recovers the incremental stress-strain relation
\begin{equation}
    \begin{aligned}
        \sig_{n+1} &= 
    \left(p_0(\theta^p_n + \Delta \theta^p) 
    + 
    \frac{\partial \psi}{\partial \dot{\theta}^p}(\frac{\Delta \theta^p}{\Delta t}, \frac{\Delta \eps^p}{\Delta t}; \theta^p_n + \Delta \theta^p) 
    + 
    \frac{\Delta \theta^p}{\Delta \eps^p}\left(\frac{1}{\tilde{\alpha}^2}
    - 
    \dfrac{2}{3}\right)\frac{\partial {\psi}}{\partial \dot{\eps}^p}(\frac{\Delta \theta^p}{\Delta t}, \frac{\Delta \eps^p}{\Delta t}; \theta^p_n + \Delta \theta^p)\right) \mathbf{I} \\
    &+ \frac{1}{3 h_{n+1}\Delta \eps^p}(\Delta \boldsymbol{\delta} \otimes \be_3 
    + 
    \be_3 \otimes \Delta \boldsymbol{\delta})\frac{\partial {\psi}}{\partial \dot{\eps}^p}(\frac{\Delta \theta^p}{\Delta t}, \frac{\Delta \eps^p}{\Delta t};\theta^p_n + \Delta \theta^p)).
    \end{aligned}
\label{eq::app_sig_del}
\end{equation}

The solution procedure for each component of~\eqref{eq::app_sig_del} differs between displacement- and load- controlled deformations. If the kinematic variables are prescribed,~\eqref{eq::app_sig_del} can be evaluated directly for the stress. For load-controlled deformations, we solve~\eqref{eq::app_sig_del} implicitly using Newton-Raphson iterations for the increment in the corresponding kinematic variables. 

In the iteration procedure, the residual is given by rearranging~\eqref{eq::app_sig_del},

\begin{equation}
    \begin{aligned}
       \mathbf{R} &= \sig - 
    \left(p_0(\theta^p_n + \Delta \theta^p) 
    + 
    \frac{\partial \psi}{\partial \dot{\theta}^p}(\frac{\Delta \theta^p}{\Delta t}, \frac{\Delta \eps^p}{\Delta t}; \theta^p_n + \Delta \theta^p)  
    + 
    \frac{\Delta \theta^p}{\Delta \eps^p}\left(\frac{1}{\tilde{\alpha}^2}
    - 
    \dfrac{2}{3}\right)\frac{\partial {\psi}}{\partial \dot{\eps}^p}(\frac{\Delta \theta^p}{\Delta t}, \frac{\Delta \eps^p}{\Delta t}; \theta^p_n + \Delta \theta^p) \right) \mathbf{I} \\
    &- \frac{1}{3 h_{n+1}\Delta \eps^p}(\Delta \boldsymbol{\delta} \otimes \be_3 
    + 
    \be_3 \otimes \Delta \boldsymbol{\delta})\frac{\partial {\psi}}{\partial \dot{\eps}^p}(\frac{\Delta \theta^p}{\Delta t}, \frac{\Delta \eps^p}{\Delta t}; \theta^p_n + \Delta \theta^p) ),
    \end{aligned}
\end{equation}
and, taking derivatives, the tangent stiffness matrix follows as
\begin{equation}
    \mathbf{K} =\frac{d \mathbf{R}}{d \Delta \boldsymbol{\delta}} = 
    \frac{\partial \mathbf{R}}{\partial \Delta \boldsymbol{\delta}}
    +
    \frac{\partial \mathbf{R}}{\partial \Delta \theta^p}\frac{\partial \Delta \theta^p}{\partial \Delta \boldsymbol{\delta}} 
    + 
    \frac{\partial \mathbf{R}}{\partial \Delta \eps^p}\frac{\partial \Delta \eps^p}{\partial \Delta \boldsymbol{\delta}}.
\end{equation}
The Newton update at iteration $(p+1)$, $\Delta \Delta \boldsymbol{\delta}^{p+1}$, is the solution of
\begin{equation}
    \mathbf{K}(\Delta \boldsymbol{\delta}^{p})\, \Delta \Delta \boldsymbol{\delta}^{p+1} = -\mathbf{R} (\Delta \boldsymbol{\delta}^{p}),
\end{equation}
and the slip increments are updated according to
\begin{equation}
    \Delta \boldsymbol{\delta}^{p+1} = \Delta \boldsymbol{\delta}^{p} + \Delta \Delta \boldsymbol{\delta}^{p+1},
\end{equation}
until the residual is smaller than a pre-defined convergence tolerance, $||\mathbf{R}(\Delta \boldsymbol{\delta}^{p+1})|| < \epsilon$. We discuss below the specific numerical implementation for the examples from Section~\ref{sec::results} of the main text.

\subsection{Oedometer test}
For the load-controlled oedometer test, $\Delta \delta_1 = \Delta \delta_2 \equiv 0$, and the normal stress $\sigma_{33} = -\sigma_{ld}(t)$ is prescribed. We first check if the stress state is on the yield surface. In this case, the dilation increment $\Delta \delta_3$ is the solution of
\begin{equation}
\begin{aligned}
    R_{33} = &-\sigma_{ld}(t_n + \Delta t) - p_0(\theta^p_n + \Delta \theta^p) 
    - 
    \dfrac{\partial {\psi}}{\partial \dot{\theta}^p}(\frac{\Delta \theta^p}{\Delta t}, \frac{|\Delta \theta^p|}{\tilde{\alpha}\Delta t}; \theta^p_n + \Delta \theta^p) \\
    &- 
    \dfrac{\operatorname{sgn}(\Delta \theta^p)}{\tilde{\alpha}} 
    \dfrac{\partial {\psi}}{\partial \dot{\eps}^p}(\frac{\Delta \theta^p}{\Delta t}, \frac{|\Delta \theta^p|}{\tilde{\alpha}\Delta t}; \theta^p_n + \Delta \theta^p),
\end{aligned}
\label{eq::app_oed_solve}
\end{equation}
where~\eqref{eq::app_oed_solve} is solved iteratively until $|R_{33}| < \epsilon$. Otherwise, e.~g. during unloading, $\Delta \delta_3 = 0$.

\subsection{Shear test}
For the shear test, we assume plane-strain conditions, $\Delta \delta_1 \equiv 0$, and prescribe the shear slip rate $\Delta \delta_2(t)$ and normal stress $\sigma_{33} = -\sigma_n$. Thus, using~\eqref{eq::discrete_eps}, the dilation increment $\Delta \delta_3$ is the solution of
\begin{equation} \label{eq::app_shear_test_resid}
    R_{33} = -\sigma_n 
    - 
    p_0(\theta^p_n + \Delta \theta^p)
    - 
    \dfrac{\partial \psi}{\partial \dot{\theta}^p}(\frac{\Delta \theta^p}{\Delta t}, \frac{\Delta \eps^p}{\Delta t}; \theta^p_n + \Delta \theta^p)
    - 
    \dfrac{\Delta{\theta}^p}{\tilde{\alpha}^2 \Delta{\eps}^p} \dfrac{\partial \psi}{\partial \dot{\eps}^p}(\frac{\Delta \theta^p}{\Delta t}, \frac{\Delta \eps^p}{\Delta t}; \theta^p_n + \Delta \theta^p)
\end{equation}
for which $|R_{33}| < \epsilon$. The shear stress, $\sigma_{23}$, follows by evaluating~\eqref{eq::app_sig_del} using the prescribed shear slip rate and the solution of~\eqref{eq::app_shear_test_resid}.

\section{Yield surface for simple shear}
\label{sec::app_sigman_hat}
We specialize the yield relation for the case of simple shear and relate the pre-consolidation stress pressure to the normal stress, $\sigma_n$, and shear stress, $\tau$. As the yield relation is written in terms of $p$ and $q$, we first rewrite~\eqref{eq::rel_yield} in terms of $\tau$ and $\sigma_n$. For plane-strain simple shear, the Cauchy stress is
\begin{equation}
    \sig = \begin{pmatrix}
        \sigma_l & 0 & 0 \\
        0 & \sigma_l & \tau \\
        0 & \tau & -\sigma_n
    \end{pmatrix}.
\end{equation}
From the definition of $q$,
\begin{equation}
    q^2 := \dfrac{3}{2}(\dev\sig)\cdot (\dev\sig) = 3 \tau^2 + \dfrac{9}{4}(\sigma_n + p)^2,
\end{equation}
and rearranging~\eqref{eq::stress_strain} in the rate-independent limit gives
\begin{equation}
    p = p_0-\dfrac{\tilde{\alpha}^2}{\alpha^2}(\sigma_n + p_0).
\end{equation}
Thus, the yield surface can be written in terms of the shear stress $\tau$ and normal stress $\sigma_n$ as
\begin{equation}
    3\tau^2 + \tilde{\alpha}\sigma_n(\tilde{\alpha}\sigma_n - 2\sigma_0) = 0,
\end{equation}
where we have used~\eqref{eq::oed_restrict}. Adopting the $\hat{\sigma}_n$ notation from the main text, this can be written as
\begin{equation}
    3\left(\dfrac{\tau}{\sigma_{c}}\right)^2 + \hat{\sigma}_n(\hat{\sigma}_n - 2\hat{\sigma}_0) = 0.
\end{equation}
The transition from dilation to compaction happens at $\hat{\sigma}_n/\so = 1$. At $t = 0$, when $\sigma_0(t=0) = \sigma_c/2$, this corresponds to $\hat{\sigma}_n = 1/2$.

\section{Additional examples for stepped shear test}
\label{sec::app_step_exs}
\begin{figure}[H]
    \centering
    \includegraphics[width=0.8\linewidth]{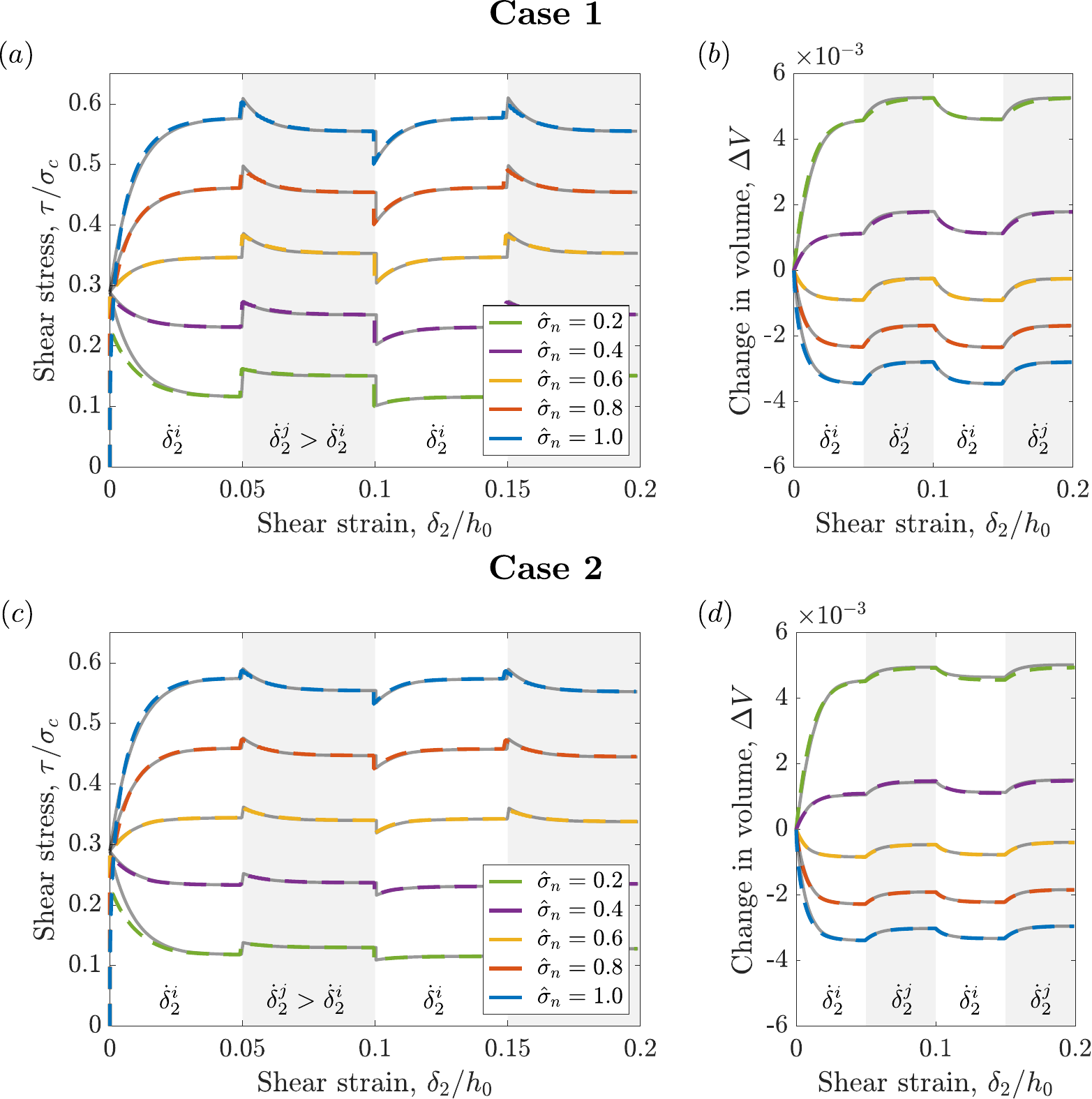}
    \caption{Numerical response to 100x stepped shear loading of a granular layer under varying normal stress, and constant consolidation pressure, for two rate-sensitivity cases. The shear displacement rate is changed every $5\%$ strain. Dashed lines correspond to numerical solutions and solid lines are theoretical predictions from~\eqref{eq::tau2} and~\eqref{eq::h_const_ld} with~\eqref{eq::sigma0_sol}. (a) and (b) Linear rate sensitivity (Case 1), with (a) shear stress and (b) volumetric strain versus shear strain. (c) and (d) Activation-motivated rate sensitivity (Case 2), with (c) shear stress and (d) volumetric strain versus shear strain.}
    \label{fig::shear_test_100x}
\end{figure}

We modify the parameters for the stepped shear tests from Section~\ref{sec::step_test} of the main text to further illustrate the model response. We first consider an example with a larger step-change in shear loading rate, with parameters shown in Table~\ref{tab::shear_params_100x}. The response of both types of rate sensitivity to the shear step-test qualitatively matches the results from the main text (Figure~\ref{fig::shear_test_100x}); there is an initial slip-rate dependent step in shear stress followed by a slip-dependent evolution to a rate-dependent steady-state. As in the example from the main text, the shear stress and volume changes for the linear case are more sensitive to rate than the activation-motivated example. 

\begin{figure}[H]
    \centering
    \includegraphics[width=0.8\linewidth]{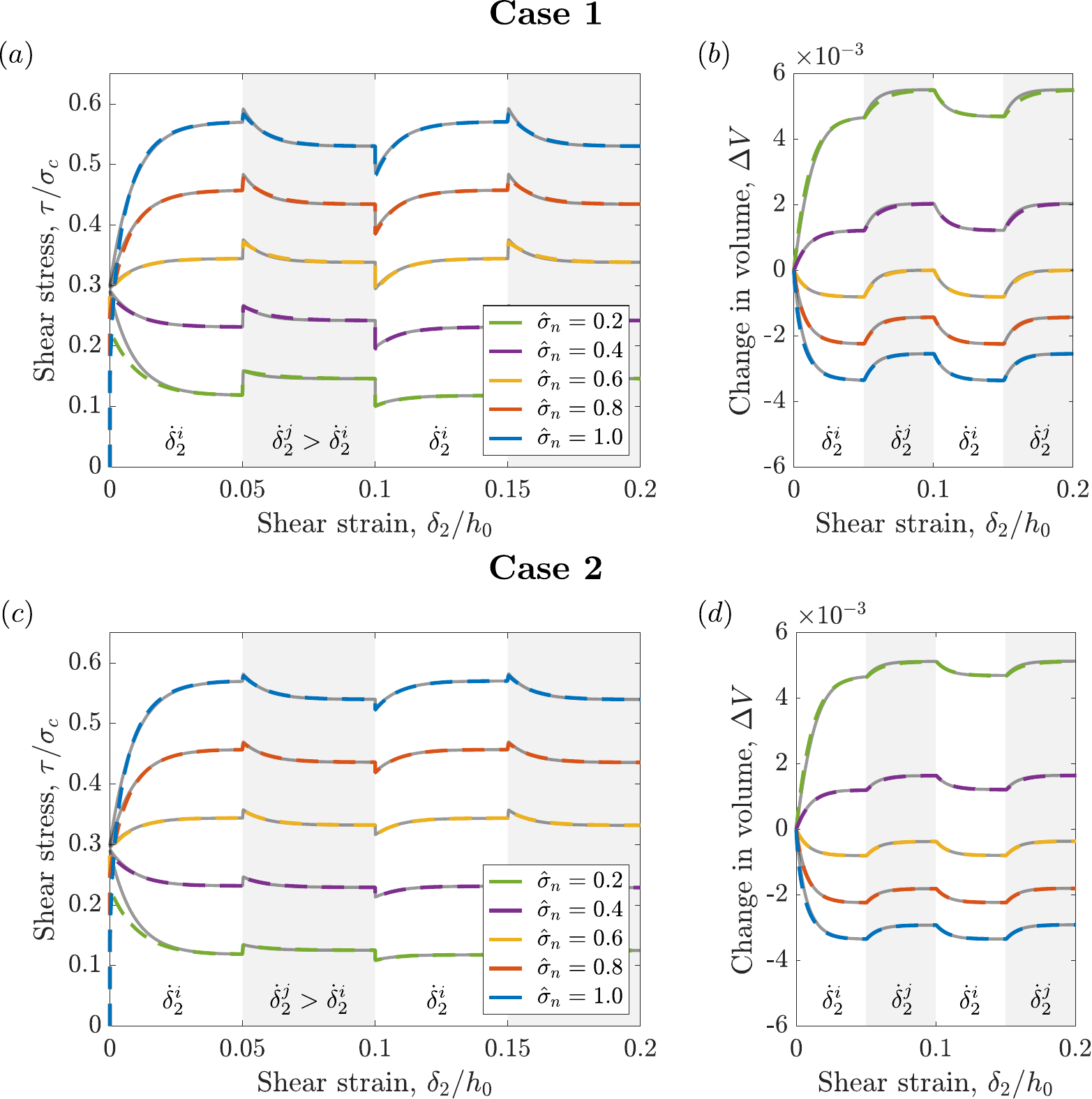}
    \caption{Numerical response to stepped shear loading of a granular layer with stronger weakening under varying normal stress, and constant consolidation pressure, for two rate-sensitivity cases. The shear displacement rate is changed every $5\%$ strain. Dashed lines correspond to numerical solutions and solid lines are theoretical predictions from~\eqref{eq::tau2} and~\eqref{eq::h_const_ld} with~\eqref{eq::sigma0_sol}. (a) and (b) Linear rate sensitivity (Case 1), with (a) shear stress and (b) volumetric strain versus shear strain. (c) and (d) Activation-motivated rate sensitivity (Case 2), with (c) shear stress and (d) volumetric strain versus shear strain.}
    \label{fig::shear_test_eta2_mod}
\end{figure}

We next modify the steady-state shear stresses by increasing the coefficient of $\tilde{\varphi}$. From~\eqref{eq::tau_ss}, this change makes the interface more steady-state rate weakening. The remaining parameters are shown in Table~\ref{tab::shear_params_eta2_mod}. The shear stress in the resulting shear step tests show more weakening for $\hat{\sigma}_n > 0.5$ than in the example considered in the main text (Figures~\ref{fig::shear_test_eta2_mod}(a) and~\ref{fig::shear_test_eta2_mod}(c)). Additionally, as predicted by~\eqref{eq::h_crit}, the larger coefficient for $\tilde{\varphi}$ results in a stronger rate dependence for the layer height (Figures~\ref{fig::shear_test_eta2_mod}(b) and~\ref{fig::shear_test_eta2_mod}(d)).

\begin{table}[H]
\begin{center}
\begin{tabular}{|p{2cm}|| p{2.5cm}|p{7cm}|} 
 \hline
 \multicolumn{3}{|c|}{\textbf{Simulation parameters}} \\
 \hline
 Symbol & Value & Description \\ 
 \hline
 $\alpha$ & $1.2$ & Triaxial internal friction coefficient \\
 $\dot{\delta}_2^i/h_0$ & $10^{-6}$ s$^{-1}$ & Initial (slow) shear loading rate \\
 $\dot{\delta}_2^j/h_0$ & $10^{-4}$ s$^{-1}$ & Fast shear loading rate \\
 $\theta^p_{\rm ref}$ & $0.005$ & Reference volumetric strain \\
 $\eta_1$ & $0.015$ & Rate-potential 1 coefficient \\
 $\eta_2$ & $0.025$ & Rate-potential 2 coefficient \\
 $\dot{\eps}^p_0$ & $10^{-5}$ s$^{-1}$ & Reference plastic strain rate \\
 \hline
\end{tabular}
\end{center}
\caption{Simulation parameters for shear test with 100x step.}
\label{tab::shear_params_100x}
\end{table}

\begin{table}[H]
\begin{center}
\begin{tabular}{|p{2cm}|| p{2.5cm}|p{7cm}|} 
 \hline
 \multicolumn{3}{|c|}{\textbf{Simulation parameters}} \\
 \hline
 Symbol & Value & Description \\ 
 \hline
 $\alpha$ & $1.2$ & Triaxial internal friction coefficient \\
 $\dot{\delta}_2^i/h_0$ & $10^{-5}$ s$^{-1}$ & Initial (slow) shear loading rate \\
 $\dot{\delta}_2^j/h_0$ & $10^{-4}$ s$^{-1}$ & Fast shear loading rate \\
 $\theta^p_{\rm ref}$ & $0.005$ & Reference volumetric strain \\
 $\eta_1$ & $0.015$ & Rate-potential $\varphi$ coefficient \\
 $\eta_2$ & $0.035$ & Rate-potential $\tilde{\varphi}$ coefficient \\
 $\dot{\eps}^p_0$ & $10^{-5}$ s$^{-1}$ & Reference plastic strain rate \\
 \hline
\end{tabular}
\end{center}
\caption{Simulation parameters for the shear test with stronger weakening.}
\label{tab::shear_params_eta2_mod}
\end{table}

\newpage
\bibliographystyle{elsarticle-harv}
\bibliography{biblio.bib}

@book{rockafellar2015convex,
  title={Convex Analysis},
  author={Rockafellar, R.T.},
  isbn={9781400873173},
  series={Princeton Landmarks in Mathematics and Physics},
  url={https://books.google.com/books?id=jzpzBwAAQBAJ},
  year={2015},
  publisher={Princeton University Press}
}

@book{lubliner2008plasticity,
  title={Plasticity Theory},
  author={Lubliner, J.},
  isbn={9780486462905},
  lccn={2007038484},
  series={Dover books on engineering},
  url={https://books.google.com/books?id=MkK-BLbHtcAC},
  year={2008},
  publisher={Dover Publications}
}

@article{ortiz2004variational,
  title={A variational Cam-clay theory of plasticity},
  author={Ortiz, M. and Pandolfi, A.},
  journal={Computer Methods in Applied Mechanics and Engineering},
  volume={193},
  number={27-29},
  pages={2645--2666},
  year={2004},
  publisher={Elsevier}
}

@article{Ortiz1999,
   author = {M. Ortiz and L. Stainier},
   doi = {10.1016/S0045-7825(98)00219-9},
   issn = {00457825},
   issue = {3-4},
   journal = {Computer Methods in Applied Mechanics and Engineering},
   title = {The variational formulation of viscoplastic constitutive updates},
   volume = {171},
   year = {1999}
}

@article{Faulkner2003,
   author = {D. R. Faulkner and A. C. Lewis and E. H. Rutter},
   doi = {10.1016/S0040-1951(03)00134-3},
   issn = {00401951},
   issue = {3-4},
   journal = {Tectonophysics},
   title = {On the internal structure and mechanics of large strike-slip fault zones: Field observations of the Carboneras fault in southeastern Spain},
   volume = {367},
   year = {2003}
}

@article{Chester1993,
   author = {F. M. Chester and J. P. Evans and R. L. Biegel},
   doi = {10.1029/92JB01866},
   issn = {01480227},
   issue = {B1},
   journal = {Journal of Geophysical Research},
   title = {Internal structure and weakening mechanisms of the San Andreas Fault},
   volume = {98},
   year = {1993}
}

@article{Byerlee1978,
   author = {J. Byerlee},
   doi = {10.1007/BF00876528},
   issn = {00334553},
   issue = {4-5},
   journal = {Pure and Applied Geophysics PAGEOPH},
   title = {Friction of rocks},
   volume = {116},
   year = {1978}
}

@article{marone1998laboratory,
  title={Laboratory-derived friction laws and their application to seismic faulting},
  author={Marone, C.},
  journal={Annual Review of Earth and Planetary Sciences},
  volume={26},
  number={1},
  pages={643--696},
  year={1998},
  publisher={Annual Reviews 4139 El Camino Way, PO Box 10139, Palo Alto, CA 94303-0139, USA},
  doi = {https://doi.org/10.1146/annurev.earth.26.1.643}
}

@article{egholm2008mechanics,
  title={The mechanics of clay smearing along faults},
  author={Egholm, D.~L. and Clausen, O.~R. and Sandiford, M. and Kristensen, M.~B. and Korstg{\aa}rd, J.~A.},
  journal={Geology},
  volume={36},
  number={10},
  pages={787--790},
  year={2008},
  publisher={Geological Society of America},
  doi = {https://doi.org/10.1130/G24975A.1}
}

@article{dunham2011earthquake,
  title={Earthquake ruptures with strongly rate-weakening friction and off-fault plasticity, part 2: Nonplanar faults},
  author={Dunham, E.~M. and Belanger, D. and Cong, L. and Kozdon, J.~E.},
  journal={Bulletin of the Seismological Society of America},
  volume={101},
  number={5},
  pages={2308--2322},
  year={2011},
  publisher={Seismological Society of America},
  doi = {https://doi.org/10.1785/0120100076}
}

@article{dolarevic2007modified,
  title={A modified three-surface elasto-plastic cap model and its numerical implementation},
  author={Dolarevic, Samir and Ibrahimbegovic, Adnan},
  journal={Computers \& structures},
  volume={85},
  number={7-8},
  pages={419--430},
  year={2007},
  publisher={Elsevier},
  doi = {https://doi.org/10.1016/j.compstruc.2006.10.001}
}

@book{scholz2019mechanics,
  title={The mechanics of earthquakes and faulting},
  author={Scholz, C.~H.},
  year={2019},
  publisher={Cambridge university press}
}

@article{mair1999friction,
  title={Friction of simulated fault gouge for a wide range of velocities and normal stresses},
  author={Mair, K. and Marone, C.},
  journal={Journal of Geophysical Research: Solid Earth},
  volume={104},
  number={B12},
  pages={28899--28914},
  year={1999},
  publisher={Wiley Online Library},
  doi = {https://doi.org/10.1029/1999JB900279}
}

@article{elbanna2014two,
  title={A two-scale model for sheared fault gouge: Competition between macroscopic disorder and local viscoplasticity},
  author={Elbanna, A.~E. and Carlson, J.~M.},
  journal={Journal of Geophysical Research: Solid Earth},
  volume={119},
  number={6},
  pages={4841--4859},
  year={2014},
  publisher={Wiley Online Library},
  doi = {https://doi.org/10.1002/2014JB011001}
}

@article{roscoe:1958,
  title={On the yielding of soils},
  author={Roscoe, K.~H. and Schofield, A.~N. and Wroth, C.~P.},
  journal={Geotechnique},
  volume={8},
  number={1},
  pages={22--53},
  year={1958},
  publisher={Thomas Telford Ltd},
  doi = {https://doi.org/10.1680/geot.1958.8.1.22}
}

@book{schofield:1968,
  title={Critical state soil mechanics},
  author={Schofield, A.~No. and Wroth, c.~P.},
  volume={310},
  year={1968},
  publisher={McGraw-hill London},
}

@article{burland:1969,
  title={Local strains and pore pressures in a normally consolidated clay layer during one-dimensional consolidation},
  author={Burland, J.~B. and Roscoe, K.~H.},
  journal={Geotechnique},
  volume={19},
  number={3},
  pages={335--356},
  year={1969},
  publisher={Thomas Telford Ltd},
  doi = {https://doi.org/10.1680/geot.1969.19.3.335}
}

@article{marone1990frictional,
  title={Frictional behavior and constitutive modeling of simulated fault gouge},
  author={Marone, C. and Raleigh, C.~B. and Scholz, C.~H.},
  journal={Journal of Geophysical Research: Solid Earth},
  volume={95},
  number={B5},
  pages={7007--7025},
  year={1990},
  publisher={Wiley Online Library},
  doi = {https://doi.org/10.1029/JB095iB05p07007}
}

@article{callari:1998,
  title={A finite-strain Cam-clay model in the framework of multiplicative elasto-plasticity},
  author={Callari, C. and Auricchio, F. and Sacco, E.},
  journal={International Journal of Plasticity},
  volume={14},
  number={12},
  pages={1155--1187},
  year={1998},
  publisher={Elsevier},
  doi = {https://doi.org/10.1016/S0749-6419(98)00050-3}
}

@article{borja:1998,
  title={Cam-Clay plasticity. Part III: Extension of the infinitesimal model to include finite strains},
  author={Borja, R.~I. and Tamagnini, C.},
  journal={Computer Methods in Applied Mechanics and Engineering},
  volume={155},
  number={1-2},
  pages={73--95},
  year={1998},
  publisher={Elsevier},
  doi = {https://doi.org/10.1016/S0045-7825(97)00141-2}
}

@article{borja:2006,
  title={Critical state plasticity. Part VI: Meso-scale finite element simulation of strain localization in discrete granular materials},
  author={Borja, R.~I and Andrade, J.~E.},
  journal={Computer Methods in Applied Mechanics and Engineering},
  volume={195},
  number={37-40},
  pages={5115--5140},
  year={2006},
  publisher={Elsevier},
  doi = {https://doi.org/10.1016/j.cma.2005.08.020}
}

@article{Rice2001,
   author = {Rice, J.~R. and Lapusta, N. and Ranjith, K.},
   issn = {00225096},
   issue = {9},
   journal = {Journal of the Mechanics and Physics of Solids},
   title = {Rate and state dependent friction and the stability of sliding between elastically deformable solids},
   volume = {49},
   year = {2001},
   doi = {https://doi.org/10.1016/S0022-5096(01)00042-4}
}

@article{cohen1959molecular,
  title={Molecular transport in liquids and glasses},
  author={Cohen, Morrel H and Turnbull, David},
  journal={The Journal of Chemical Physics},
  volume={31},
  number={5},
  pages={1164--1169},
  year={1959},
  publisher={American Institute of Physics},
  doi = {https://doi.org/10.1063/1.1730566}
}

@article{edwards1989theory,
  title={Theory of powders},
  author={Edwards, S.~F. and Oakeshott, R.~B.~S.},
  journal={Physica A: Statistical Mechanics and its Applications},
  volume={157},
  number={3},
  pages={1080--1090},
  year={1989},
  publisher={Elsevier},
  doi = {https://doi.org/10.1016/0378-4371(89)90034-4}
}

@article{Brzesowsky2014a,
   author = {R. H. Brzesowsky and C. J. Spiers and C. J. Peach and S. J.T. Hangx},
   doi = {10.1002/2013JB010444},
   issn = {21699356},
   issue = {2},
   journal = {Journal of Geophysical Research: Solid Earth},
   title = {Time-independent compaction behavior of quartz sands},
   volume = {119},
   year = {2014}
}

@book{Das2019,
   author = {Braja M. Das},
   doi = {10.1201/9781351215183},
   journal = {Advanced Soil Mechanics},
   title = {Advanced Soil Mechanics},
   year = {2019}
}

@article{Rathbun2013,
   author = {Andrew P. Rathbun and Chris Marone},
   doi = {10.1002/jgrb.50224},
   issn = {21699356},
   issue = {7},
   journal = {Journal of Geophysical Research: Solid Earth},
   title = {Symmetry and the critical slip distance in rate and state friction laws},
   volume = {118},
   year = {2013}
}

@article{Leroueil1985,
   author = {S. Leroueil and M. Kabbaj and F. Tavenas and R. Bouchard},
   doi = {10.1680/geot.1985.35.2.159},
   issn = {17517656},
   issue = {2},
   journal = {Geotechnique},
   title = {Stress-strain-strain rate relation for the compressibility of sensitive natural clays},
   volume = {35},
   year = {1985}
}

@article{Dieterich1978,
   author = {James H. Dieterich},
   doi = {10.1007/BF00876539},
   issn = {00334553},
   issue = {4-5},
   journal = {Pure and Applied Geophysics PAGEOPH},
   title = {Time-dependent friction and the mechanics of stick-slip},
   volume = {116},
   year = {1978}
}

@article{Rabinowicz1958,
   author = {Ernest Rabinowicz},
   doi = {10.1088/0370-1328/71/4/316},
   issn = {03701328},
   issue = {4},
   journal = {Proceedings of the Physical Society},
   title = {The intrinsic variables affecting the stick-slip process},
   volume = {71},
   year = {1958}
}

@article{Rabinowicz1951,
   author = {Ernest Rabinowicz},
   doi = {10.1063/1.1699869},
   issn = {00218979},
   issue = {11},
   journal = {Journal of Applied Physics},
   title = {The nature of the static and kinetic coefficients of friction},
   volume = {22},
   year = {1951}
}

@article{Karig2003,
   author = {D. E. Karig and M. V.S. Ask},
   doi = {10.1029/2001jb000652},
   issn = {21699356},
   issue = {4},
   journal = {Journal of Geophysical Research: Solid Earth},
   title = {Geological perspectives on consolidation of clay-rich marine sediments},
   volume = {108},
   year = {2003}
}

@article{Choens2018,
   author = {R. C. Choens and F. M. Chester},
   doi = {10.1029/2017JB015097},
   issn = {21699356},
   issue = {8},
   journal = {Journal of Geophysical Research: Solid Earth},
   title = {Time-Dependent Consolidation in Porous Geomaterials at In Situ Conditions of Temperature and Pressure},
   volume = {123},
   year = {2018}
}

@article{Bjerrum1967,
   author = {Laurits Bjerrum},
   doi = {10.1680/geot.1967.17.2.83},
   issn = {17517656},
   issue = {2},
   journal = {Geotechnique},
   title = {Engineering geology of norwegian normally-consolidated marine clays as related to settlements of buildings},
   volume = {17},
   year = {1967}
}

@inproceedings{Suklje1957,
   author = {L. Šuklje},
   booktitle = {4th International Conference on Soil Mechanics and Foundation Engineering},
   title = {The Analysis of the Consolidation Process by the Isotaches Method},
   year = {1957}
}

@article{Laloui2008,
   author = {L. Laloui and S. Leroueil and S. Chalindar},
   doi = {10.1139/T08-093},
   issn = {00083674},
   issue = {12},
   journal = {Canadian Geotechnical Journal},
   title = {Modelling the combined effect of strain rate and temperature on one-dimensional compression of soils},
   volume = {45},
   year = {2008}
}

@article{Segall1995,
   author = {P. Segall and J. R. Rice},
   doi = {10.1029/95jb02403},
   issn = {01480227},
   issue = {B11},
   journal = {Journal of Geophysical Research},
   title = {Dilatancy, compaction, and slip instability of a fluid-infiltrated fault},
   volume = {100},
   year = {1995}
}

@article{Sleep1995,
   author = {N. H. Sleep},
   doi = {10.1029/94jb03340},
   issn = {01480227},
   issue = {B7},
   journal = {Journal of Geophysical Research},
   title = {Ductile creep, compaction, and rate and state dependent friction within major fault zones},
   volume = {100},
   year = {1995}
}

@article{Rubino2017,
   author = {V. Rubino and A. J. Rosakis and N. Lapusta},
   doi = {10.1038/ncomms15991},
   issn = {20411723},
   journal = {Nature Communications},
   title = {Understanding dynamic friction through spontaneously evolving laboratory earthquakes},
   volume = {8},
   year = {2017},
}

@article{Dieterich1979,
   author = {James H. Dieterich},
   doi = {10.1029/JB084iB05p02161},
   issn = {21699356},
   issue = {B5},
   journal = {Journal of Geophysical Research: Solid Earth},
   title = {Modeling of rock friction: 1. {Experimental} results and constitutive equations},
   volume = {84},
   year = {1979},
}

@article{Ruina1983,
   author = {A. Ruina},
   doi = {10.1029/JB088iB12p10359},
   issn = {01480227},
   issue = {B12},
   journal = {Journal of Geophysical Research},
   title = {Slip instability and state variable friction laws.},
   volume = {88},
   year = {1983},
}

@article{Rice1983,
   author = {J. R. Rice and A. L. Ruina},
   doi = {10.1115/1.3167042},
   issn = {15289036},
   issue = {2},
   journal = {Journal of Applied Mechanics, Transactions ASME},
   title = {Stability of steady frictional slipping},
   volume = {50},
   year = {1983},
}

@article{Lee1964,
   author = {E. H. Lee},
   doi = {10.1115/1.3564580},
   issn = {15289036},
   issue = {1},
   journal = {Journal of Applied Mechanics, Transactions ASME},
   title = {Elastic-plastic deformation at finite strains},
   volume = {36},
   year = {1964}
}

@article{Chester1998,
   author = {Frederick M. Chester and Judith S. Chester},
   doi = {10.1016/S0040-1951(98)00121-8},
   issn = {00401951},
   issue = {1-2},
   journal = {Tectonophysics},
   title = {Ultracataclasite structure and friction processes of the Punchbowl fault, San Andreas system, California},
   volume = {295},
   year = {1998}
}

@article{Barbot2022,
   author = {Sylvain Barbot},
   doi = {10.1029/2022JB025106},
   issn = {21699356},
   issue = {11},
   journal = {Journal of Geophysical Research: Solid Earth},
   title = {A Rate-, State-, and Temperature-Dependent Friction Law With Competing Healing Mechanisms},
   volume = {127},
   year = {2022}
}

@article{Niemeijer2007,
   author = {A. R. Niemeijer and C. J. Spiers},
   doi = {10.1029/2007JB005008},
   issn = {21699356},
   issue = {10},
   journal = {Journal of Geophysical Research: Solid Earth},
   title = {A microphysical model for strong velocity weakening in phyllosilicate-bearing fault gouges},
   volume = {112},
   year = {2007}
}

@article{Collins-Craft2020,
   author = {Nicholas Anton Collins-Craft and Ioannis Stefanou and Jean Sulem and Itai Einav},
   doi = {10.1016/j.jmps.2020.103975},
   issn = {00225096},
   journal = {Journal of the Mechanics and Physics of Solids},
   title = {A Cosserat Breakage Mechanics model for brittle granular media},
   volume = {141},
   year = {2020}
}

@article{Henann2013,
   author = {David L. Henann and Ken Kamrin},
   doi = {10.1073/pnas.1219153110},
   issn = {00278424},
   issue = {17},
   journal = {Proceedings of the National Academy of Sciences of the United States of America},
   title = {A predictive, size-dependent continuum model for dense granular flows},
   volume = {110},
   year = {2013}
}

@misc{Vakis2018,
   author = {A. I. Vakis and V. A. Yastrebov and J. Scheibert and L. Nicola and D. Dini and C. Minfray and A. Almqvist and M. Paggi and S. Lee and G. Limbert and J. F. Molinari and G. Anciaux and R. Aghababaei and S. Echeverri Restrepo and A. Papangelo and A. Cammarata and P. Nicolini and C. Putignano and G. Carbone and S. Stupkiewicz and J. Lengiewicz and G. Costagliola and F. Bosia and R. Guarino and N. M. Pugno and M. H. Müser and M. Ciavarella},
   doi = {10.1016/j.triboint.2018.02.005},
   issn = {0301679X},
   journal = {Tribology International},
   title = {Modeling and simulation in tribology across scales: An overview},
   volume = {125},
   year = {2018}
}

@article{Nagy2017,
   author = {Dániel B. Nagy and Philippe Claudin and Tamás Börzsönyi and Ellák Somfai},
   doi = {10.1103/PhysRevE.96.062903},
   issn = {24700053},
   issue = {6},
   journal = {Physical Review E},
   title = {Rheology of dense granular flows for elongated particles},
   volume = {96},
   year = {2017}
}

@article{Bilotto2025,
	author = {Bilotto, J. and Trulsson, M. and Molinari, J.-F.},
	title = {Numerical study of simple shear dense granular flow of frictional elongated and flattened particles},
	DOI= "10.1051/epjconf/202534002008",
	url= "https://doi.org/10.1051/epjconf/202534002008",
	journal = {EPJ Web Conf.},
	year = 2025,
	volume = 340,
	pages = "02008",
}

@article{Tsutsumi1997,
   author = {Akito Tsutsumi and Toshihiko Shimamoto},
   doi = {10.1029/97GL00503},
   issn = {00948276},
   issue = {6},
   journal = {Geophysical Research Letters},
   title = {High-velocity frictional properties of gabbro},
   volume = {24},
   year = {1997}
}

@article{DiToro2011,
   author = {G. {Di Toro} and R. Han and T. Hirose and N. De Paola and S. Nielsen and K. Mizoguchi and F. Ferri and M. Cocco and T. Shimamoto},
   doi = {10.1038/nature09838},
   issn = {00280836},
   issue = {7339},
   journal = {Nature},
   title = {Fault lubrication during earthquakes},
   volume = {471},
   year = {2011}
}

@article{Lapusta2000,
   author = {Nadia Lapusta and James R. Rice and Yehuda Ben-Zion and Gutuan Zheng},
   doi = {10.1029/2000jb900250},
   issn = {21699356},
   issue = {B10},
   journal = {Journal of Geophysical Research: Solid Earth},
   title = {Elastodynamic analysis for slow tectonic loading with spontaneous rupture episodes on faults with rate- and state-dependent friction},
   volume = {105},
   year = {2000},
}

@article{Rice2006,
   author = {James R. Rice},
   doi = {10.1029/2005JB004006},
   issn = {21699356},
   issue = {5},
   journal = {Journal of Geophysical Research: Solid Earth},
   title = {Heating and weakening of faults during earthquake slip},
   volume = {111},
   year = {2006}
}

@inbook{Tullis2015,
   author = {T. E. Tullis},
   doi = {10.1016/B978-0-444-53802-4.00073-7},
   booktitle = {Treatise on Geophysics: Second Edition},
   title = {Mechanisms for Friction of Rock at Earthquake Slip Rates},
   volume = {4},
   year = {2015}
}

@article{Hulikal2015,
   author = {Srivatsan Hulikal and Kaushik Bhattacharya and Nadia Lapusta},
   doi = {10.1016/j.jmps.2014.10.008},
   issn = {00225096},
   journal = {Journal of the Mechanics and Physics of Solids},
   title = {Collective behavior of viscoelastic asperities as a model for static and kinetic friction},
   volume = {76},
   year = {2015}
}

@article{Tengattini2016,
   author = {A. Tengattini and A. Das and I. Einav},
   doi = {10.1680/jgeot.14.P.164},
   issn = {17517656},
   issue = {9},
   journal = {Geotechnique},
   title = {A constitutive modelling framework predicting critical state in sand undergoing crushing and dilation},
   volume = {66},
   year = {2016}
}

@article{Einav2007,
   author = {Itai Einav},
   doi = {10.1016/j.jmps.2006.11.003},
   issn = {00225096},
   issue = {6},
   journal = {Journal of the Mechanics and Physics of Solids},
   title = {Breakage mechanics-Part I: Theory},
   volume = {55},
   year = {2007}
}

@article{Bedford2021,
   author = {John D. Bedford and Daniel R. Faulkner},
   doi = {10.1029/2020GL092023},
   issn = {19448007},
   issue = {7},
   journal = {Geophysical Research Letters},
   title = {The Role of Grain Size and Effective Normal Stress on Localization and the Frictional Stability of Simulated Quartz Gouge},
   volume = {48},
   year = {2021}
}

@book{Borja2013,
   author = {Ronaldo I. Borja},
   issue = {3},
   journal = {Books},
   title = {Plasticity: Modeling \& Computation},
   volume = {4},
   year = {2013}
}

@article{Bedford2022,
   author = {John D. Bedford and Daniel R. Faulkner and Nadia Lapusta},
   doi = {10.1038/s41467-022-27998-2},
   issn = {20411723},
   issue = {1},
   journal = {Nature Communications},
   title = {Fault rock heterogeneity can produce fault weakness and reduce fault stability},
   volume = {13},
   year = {2022}
}

@article{Beeler1994,
   author = {N. M. Beeler and T. E. Tullis and J. D. Weeks},
   doi = {10.1029/94GL01599},
   issn = {19448007},
   issue = {18},
   journal = {Geophysical Research Letters},
   title = {The roles of time and displacement in the evolution effect in rock friction},
   volume = {21},
   year = {1994}
}

@article{Beeler1996,
   author = {N. M. Beeler and T. E. Tullis and M. L. Blanpied and J. D. Weeks},
   doi = {10.1029/96jb00411},
   issn = {21699356},
   issue = {4},
   journal = {Journal of Geophysical Research: Solid Earth},
   title = {Frictional behavior of large displacement experimental faults},
   volume = {101},
   year = {1996}
}

@incollection{roscoe:1968,
  author         = {Roscoe, K.~H. and Burland, J.~B.},
  booktitle      = {Engineering Plasticity},
  title          = {On the generalised stress-strain behaviour of ``wet'' clay.},
  publisher      = {Cambridge University Press},
  year           = 1968,
  editors        = {Heyman, J. and Leckie, F.~A.},
  address        = {Cambridge}
}

\end{document}